\documentclass[final,5p,times,twocolumn]{elsarticle}
 \biboptions{comma,sort&compress}
\usepackage{graphicx}

\usepackage{ecrc}

\volume{00}

\firstpage{1}

\journalname{Nuclear and Particle Physics Proceedings}

\runauth{}

\jid{nppp}

\jnltitlelogo{Nuclear and Particle Physics Proceedings}

\usepackage{amssymb}

\usepackage[figuresright]{rotating}

\def\beq{\begin{equation}}
\def\eeq{\end{equation}}
\def\bea{\begin{eqnarray}}
\def\eea{\end{eqnarray}}
\def\bq{\begin{quote}}
\def\eq{\end{quote}}

\def\nnb{\nonumber}
\def\ga{\left(}
\def\dr{\right)}

\def\nnb{\nonumber}
\def\la{\langle}
\def\ra{\rangle}
\def\nin{\noindent}
\def\ba{\vspace*{-0.2cm}\begin{array}}
\def\ea{\end{array}\vspace*{-0.2cm}}

\def\b{$\bullet~$}
\def\als{\alpha_s}

\def\gg2{ \la\alpha_s G^2 \ra}
\def\gg3{g^3f_{abc}\la G^aG^bG^c \ra}
\def\ggg4{\la\als^2G^4\ra}

\def\beq{\begin{equation}}
\def\enq{\end{equation}}
\def\beqa{\begin{eqnarray}}
\def\enqa{\end{eqnarray}}
\def\nnb{\nonumber}

\def\qq{\lag\bar{q}q\rag}

\def\lb{\label}

\newcommand{\rag}{\rangle}
\newcommand{\lag}{\langle}

\def\tbl{\caption}

\journal{Elsevier}

\begin{document}

\begin{frontmatter}

\title{Decay Constants of Heavy-Light Mesons  from QCD\tnoteref{cor0}\tnoteref{cor1}
}
\tnotetext[cor0]{Plenary talk given at QCD15 (18th International QCD Conference  and 30th Anniversary), 29 june-3 july 2015 (Montpellier-FRA) and at HEPMAD15 (7th International Conference), 17-22 september 2015 (Antananarivo-MGA). 
}
\tnotetext[cor1]{\it En hommage aux victimes des attentats du 13 Novembre 2015 \`a Paris-FRA.}
 \author[label2]{Stephan Narison}
\address[label2]{Laboratoire
Particules et Univers de Montpellier, CNRS-IN2P3, 
Case 070, Place Eug\`ene
Bataillon, 34095 - Montpellier, France.}
   \ead{snarison@yahoo.fr}
\begin{abstract}
\nin
We summarize our recently improved results  for the pseudoscalar \cite{SNFB12a,SNFB12b}, vector and $B_c$\,\cite{SNFB15} meson decay constants  from QCD spectral sum rules where N2LO $\oplus$ estimate of the N3LO PT $\oplus ~d\leq 6 $ condensates have been included in the SVZ expansion. The ``optimal results"   
based on stability criteria with respect to the variations of the Laplace/Moments sum rule variables, QCD continuum
threshold and subtraction constant $\mu$ are compared with recent sum rules and lattice calculations.  To understand the ``apparent tension" between some recent results, we present  in Section 8 {\it a novel extraction of for $f_{B^*}/f_B$ from heavy quark effective theory (HQET) sum rules} by including the normalization factor $(M_b/M_B)^2$ relating the pseudoscalar to the universal HQET correlators 
for finite $b$-quark and $B$-meson masses.
 We obtain $f_{B^*}/f_B=1.025(16)$ in good agreement with  1.016(16) from spectral sum rules (SR) in full QCD \cite{SNFB15}. We complete the paper by including  {\it new improved estimates of the scalar, axial and  $B^*_c$ meson decay constants} (Sections 11--13).  For further phenomenological uses, we attempt to extract a Global Average of the sum rules and lattice determinations 
 which are summarized in Tables \ref{tab:fp}--\ref{tab:fbc}. We do not  found any deviation of these SM results from the present data. 
\end{abstract}
\end{frontmatter}
\section{Introduction}
\vspace*{-0.2cm}
\nin
The decay constants $f_{P/S}, f_{V/A}$ of (pseudo)scalar and (axial)vector mesons are of prime interests for understanding the realizations of chiral symmetry in QCD and for controlling the meson (semi-)leptonic decay widths, hadronic couplings and form factors\,\footnote{Some applications to $D$ and $B$-decays can e.g be found in \cite{GRINSTEIN,BECIR2,ROSNER,MAHMOUDI,SNFDTEST}.}. 
In addition to the well-known values of $f_\pi$=130.4(2) MeV  and $f_K$=156.1(9) MeV which control  the light flavour chiral symmetries \cite{ROSNER}, it is also desirable to extract the ones of the heavy-light charm and bottom quark systems with high-accuracy. 
This improvement program has been initiated by the recent predictions of $f_{D_{(s)}},~f_{B_{(s)}}$ \cite{SNFB12a,SNFB12b} and pursued for $f_{D^*_{(s)}},~f_{B^*_{(s)}}$ and $f_{B_c}$ in \cite{SNFB15} from QCD spectral sum rules (QSSR) \cite{SVZ} which are improved predictions of earlier estimates \cite{SNZ,GENERALIS,SNFB,SNFB88,SNFB3,SNFBREV,SNhl,
JAMIN3a,JAMIN3b,JAMIN3c,NEUBERT2,BALL,PENIN}\,\footnote{For reviews with complete references, see e.g: \cite{SNB1,SNB2,SNB3,RRY,CK,SNFB12a,SNFB12b,SNFBREV}.}
since the pioneering work of Novikov et al. (NSV2Z) \cite{NSV2Z}. Here, these decay constants are normalized as:
\beq
\hspace*{-0.6cm}\la 0|J_{P/S}(x)|P\ra= f_{P/S} M_{P/S}^2; ~
\la 0|J_{V/A}^\mu(x)|V/A\ra= f_{V/A} M_V{V/A}\epsilon^\mu,
\label{eq:fp}
\label{eq:current}
\eeq
where: 
$\epsilon_\mu$ is the vector polarization; $J_{P/S}(x)\equiv (m_q+/- M_Q) \bar q(i\gamma_5/1)
Q$ 
and    $J_{V/A}^\mu(x)\equiv \bar q(\gamma_\mu/ \gamma_\mu \gamma_5) Q$ 
are the local heavy-light pseudoscalar, scalar and vector currents;  $q\equiv d,c;~Q\equiv c,b;~P\equiv D,B,B_c$; $S\equiv D^*_0, B^*_0$, $V\equiv D^*,B^*$, $A\equiv D_1,B_1$ and their  $s$-quark analogue; the decay constants $f_{P/S}, f_{V/A}$  are related to  the leptonic widths $\Gamma [P/S\to l^+\nu_l]$ and $\Gamma [V/A\to l^+l^-]$.
The associated  two-point correlators are:
\vspace*{-0.25cm}
 \bea
\psi_{P/S}(q^2)&=&i\int d^4x ~e^{iq.x}\lag 0|TJ_{P/S}(x)J_{P/S}(0)^\dagger|0\rag~,\nnb\\
\Pi_{V/A}^{\mu\nu}(q^2)&=&i\int d^4x ~e^{iq.x}\lag 0
|TJ_{V/A}^{\mu}(x)J_{V/A}^{\nu}(0)^\dagger|0\rag~\nnb\\
&=&-\ga g^{\mu\nu}-{q^\mu q^\nu\over q^2}\dr\Pi^T_{V/A}(q^2)+{q_\mu q_\nu\over q^2} \Pi^L_{V/A}(q^2)
\lb{2po}
\eea
where $\Pi^T_{V/A}(q^2)$ has  more power of $q^2$ than the   transverse two-point function used in the current literature for $m_q=m_Q$ in order to avoid mass singularities at $q^2=0$ when $m_q$ goes to zero. $\Pi^{\mu\nu}_{V/A}(q^2)$ (vector/axial) and $\psi^{S/P}(q^2)$ (scalar/pseudoscalar) correlators are related each other through the Ward identities:
\vspace*{-0.25cm}
\beq
q_\mu q_\nu\Pi^{\mu\nu}_{V/A}(q^2)= \psi_{S/P}(q^2)-\psi_{S/P}(0)~,
 \eeq
 \vspace*{-0.25cm}
 where the (non)perturbative parts of $\psi_{S/P}(0)$ are known (see e.g. \cite{SNB1})
 and play an important r\^ole  for absorbing mass singularities which appear during the  evaluation of the PT two-point function.
For extracting the decay constants, we shall work with the well established (inverse) Laplace sum rules
{
[we use the terminology : (inverse) Laplace instead of Borel sum rule as it has been demonstrated in \cite{SNRAF} that its QCD radiative corrections  satisfy the (inverse) Laplace transform properties]}\,\footnote{One can also work with moment sum rules:
\beq
{\cal M}^{(n)}_{\bar qb}(\mu)=\int_{(m_q+M_b)^2}^{t_c}{dt\over t^{n+2}}~\frac{1}{\pi} \mbox{Im}\psi_{\bar qb}(t,\mu)~,
\label{eq:mom}
\eeq
 like in \cite{SNFB12a} or with $\tau$-decay like finite energy sum rules \cite{SNmass} inspired from $\tau$-decay \cite{BNP} but these different sum rules give approximately the same results as the one from (inverse) Laplace sum rules.} :
\beq
{\cal L}_{P/S,V/A}(\tau,\mu)=\int_{(m_q+M_Q)^2}^{t_c}dt~e^{-t\tau}\frac{1}{\pi} \mbox{Im}~\big{[}\psi_{P/S},\Pi^T_{V/A}\big{]}(t,\mu)~.
\label{eq:lsr}
\eeq
For improving the extraction of the decay constants $f_{D^*_{(s)}}$ and $f_{B^*_{(s)}}$, we shall also work with the  ratio:
\beq
{\cal R}_{V/A,P}\equiv {{\cal L}_{V,A}(\tau_{V/A})\over {\cal L}_P(\tau_P)}~,
\label{eq:ratio}
\eeq
 in order to minimize  the systematics of the approach, the effects of heavy quark masses and the continuum threshold uncertainties which are one of the main sources of errors in the determinations of the individual decay constants. $\tau_{V/A}, \tau_{P}$ denote the values of the sum rule $\tau$-variable at which each individual sum rule is optimized (minimum or inflexion point). In general,  $\tau_{V/A}\not= \tau_{P}$ as we shall see later on which requires some care for a precise determination of the ratio of decay constants. This ratio of sum rule has lead to a successful prediction of the SU(3) breaking ratio $f_{P_s}/f_{P}$ \cite{SNFBSU3} such that, from it, one expects to extract   precise values of the ratio $f_{V/A}/f_{P}$. 
\section{QCD expression of the two-point correlators}
\nin
The QCD expression of the Laplace sum rule ${\cal L}_{P/S}(\tau,\mu)$ in the (pseudo)scalar channel can be found in \cite{SNFB12a,SNFB12b,SNhl} for full QCD including N2LO perturbative QCD corrections and contributions of non-perturbative condensates up to the complete $d=6$ dimension condensates\,\footnote{Note an unfortunate missprint of $1/\pi$ in front of $\mbox{Im}\psi(t)$ in Ref. \cite{SNFB12a}.}. The one of the vector channel has been given within the same approximation in \cite{SNFB15}. Some comments are in order :

\b The expressions of NLO PT in \cite{BROAD,BROAD1}, of N2LO PT  in  \cite{CHET}, of the non-perturbative in \cite{NSV2Z,GENERALIS} and the light quark mass corrections in \cite{BROAD,GENERALIS,JAMIN} have been used. The N3LO PT contributions have been estimated by assuming the  geometric growth of the series \cite{NZ} which is dual to the effect of a $1/q^2$ term \cite{CNZ,ZAK}. The PT expressions have been originally obtained using a pole quark mass which is transformed into the $\overline{MS}$-scheme by using the relation between the running $\bar{m}_Q(\mu)$ and pole mass
$M_Q$ in the $\overline{MS}$-scheme to order $\alpha_s^2$ \cite{TAR,COQUE,SNPOLE,BROAD2,CHET2}.

\b The LO contributions in $\alpha_s$ up to the $d=4$ gluon: 
$\la\alpha_sG^2\ra\equiv \la \alpha_s G^a_{\mu\nu}G_a^{\mu\nu}\ra~,
$  $d=5$ mixed quark-gluon: 
$\la\bar qGq\ra\equiv {\la\bar qg\sigma^{\mu\nu} (\lambda_a/2) G^a_{\mu\nu}q\ra}=M_0^2\la \bar qq\ra
$
 and $d=6$ quark:
$\la\bar dj d\ra\equiv\la \bar d g\gamma_\mu D^\mu G_{\mu\nu}{\lambda_a\over 2} d\ra=
g^2\la \bar d \gamma_\mu {\lambda_a\over 2} d\sum_q \bar q \gamma_\mu {\lambda_a\over 2} q\ra
\simeq-{16\over 9} (\pi\alpha_s)~\rho\la \bar dd\ra^2,
$
(after the use of the equation of motion) condensates,  have been obtained originally by NSV2Z~\cite{NSV2Z}~. $\rho \simeq 3-4$ measures the deviation from the vacuum saturation estimate of the $d=6$ four-quark condensates \cite{SNTAU,LNT,JAMI2}.

\b The contribution of the  $d=6$ gluon condensates:
$
\la g^3G^3\ra\equiv \la g^3f_{abc}G^a_{\mu\nu}G^b_{\nu\rho}G^c_{\rho\mu}\ra~,
 \la j^2\ra\equiv g^2 \la (D_\mu G^a_{\nu\mu})^2\ra
=g^4\la\ga \sum_q\bar q\gamma_\nu {\lambda^a\over 2}q\dr^2\hspace*{-0.1cm}\ra 
\simeq -{64\over 3} (\pi\alpha_s)^2 \rho\la\bar dd\ra^2,
$
after the use of the equation of motion, 
have been deduced from \cite{GENERALIS} (Eqs. II.4.28 and Table II.8). 

\b  One can notice that the gluon condensate $\la\alpha_sG^2\ra$ and $\la G^3\ra$ contributions flip sign from the pseudoscalar 
to the vector correlators while there is an extra $m_Q M_0^2\la \bar dd\ra$ term with a positive contribution in the pseudoscalar channel from the mixed condensate. Similar observation arises for the signs of the chiral quark and mixed quark gluon condensates from the pseudoscalar to the scalar and from the vector to the axial correlators. We shall see in Fig. \ref{fig:fdtau} that these different signs transform e.g the minimum in $\tau$ for the pseudoscalar into an inflexion point for the vector sum rule.  

\b It is clear that, for some non-perturbative terms which are known to leading order
of perturbation theory, one can use either the running or the pole 
mass. However, we shall see that this distinction does not affect, in a visible way, the present result, within the accuracy of our estimate, as the non-perturbative contributions are relatively small though vital in the analysis.
{\scriptsize
\begin{table}[hbt]
 \caption{QCD input parameters:
the original errors for 
$\la\alpha_s G^2\ra$, $\la g^3  G^3\ra$ and $\rho \la \bar qq\ra^2$ have been multiplied by 2-3 for a conservative estimate. }  
\setlength{\tabcolsep}{1.1pc}
    {\scriptsize
 \begin{tabular}{lll}
&\\
\hline
\hline
Parameters&Values& Ref.    \\
\hline
$\alpha_s(M_\tau)$& $0.325(8)$&\cite{SNTAU,BNP,BETHKE,PDG}\\
$\overline{m}_c(m_c)$&$1261(12)$ MeV &average \cite{SNH10,PDG,IOFFE}\\
$\overline{m}_b(m_b)$&$4177(11)$ MeV&average \cite{SNH10,PDG}\\
$\hat \mu_q$&$(253\pm 6)$ MeV&\cite{SNB1,SNmass,SNmass98,SNLIGHT}\\
$M_0^2$&$(0.8 \pm 0.2)$ GeV$^2$&\cite{JAMI2,HEID,SNhl}\\
$\la\alpha_s G^2\ra$& $(7\pm 3)\times 10^{-2}$ GeV$^4$&
\cite{SNTAU,LNT,SNI,YNDU,SNHeavy,BELL,SNH10,SNG1,SNGH}\\
$\la g^3  G^3\ra$& $(8.2\pm 2.0)$ GeV$^2\times\la\alpha_s G^2\ra$&
\cite{SNH10}\\
$\rho \alpha_s\la \bar qq\ra^2$&$(5.8\pm 1.8)\times 10^{-4}$ GeV$^6$&\cite{SNTAU,LNT,JAMI2}\\
$\hat m_s$&$(0.114\pm0.006)$ GeV &\cite{SNB1,SNTAU9,SNmass,SNmass98,SNLIGHT}\\
$\kappa\equiv \la \bar ss\ra/\la\bar dd\ra$& $(0.74\pm  0.12)$&\cite{HBARYON,SNB1}\\
\hline\hline
\label{tab:param}
\end{tabular}
}
\end{table}
} 
\vspace*{-0.5cm}
\section{QCD input parameters}
\nin
The QCD parameters which shall appear in the following analysis will be the charm and bottom quark masses $m_{c,b}$ (we shall neglect  the light quark masses $q\equiv u,d$),
the light quark condensate $\qq$,  the gluon condensates $ \lag
\alpha_sG^2\rag$
and $ \la g^3G^3\ra$
the mixed condensate $\la\bar qGq\ra$ defined previously
and the four-quark 
 condensate $\rho\alpha_s\la\bar qq\ra^2$.
Their values are given in Table \ref{tab:param}. 

\b We shall work with the running light quark condensates and masses. 
They read:
\bea
{\la\bar qq\ra}(\tau)&=&-{\hat \mu_q^3  \ga-\beta_1a_s\dr^{2/{
\beta_1}}}/C(a_s)
\nnb\\
{\la\bar q Gq\ra}(\tau)&=&-{M_0^2{\hat \mu_q^3} \ga-\beta_1a_s\dr^{1/{3\beta_1}}}/C(a_s)~,\nnb\\
\overline{m}_s(\tau)&=&{\hat m_s/{\ga -{\rm Log}{\sqrt{\tau}\Lambda}\dr ^{2/-\beta_1}}}C(a_s)~,
\label{eq:run}
\eea
where $\beta_1=-(1/2)(11-2n_f/3)$ is the first coefficient of the $\beta$ function 
for $n_f$ flavours; $a_s\equiv \alpha_s(\tau)/\pi$; 
$\hat\mu_q$ is the spontaneous RGI light quark condensate \cite{FNR}. The QCD correction factor $C(a_s)$ in the previous expressions is numerically \cite{CHET2}:
\bea
C(a_s)=1+1.1755a_s+1.5008a_s^2 +...~~{\rm :}~~ n_f=5~,
\eea
which shows a good convergence. 

\b The value of the running $\la \bar qq\ra$ condensate is deduced from  the well-known GMOR relation: 
$
(m_u+m_d)\la \bar uu+\bar dd\ra=-m_\pi^2f_\pi^2~,
$
where $f_\pi=130.4(2)$ MeV \cite{ROSNER} and the value of $(\overline{m}_u+\overline{m}_d)(2)=(7.9\pm 0.6)$ MeV obtained in  \cite{SNmass} which agrees with the PDG  in \cite{PDG}  and lattice averages in \cite{LATT13}. Then, we deduce the RGI light quark spontaneous mass $\hat\mu_q$ given  in Table~\ref{tab:param}. 

\b For the heavy quarks, we shall use the running mass and the corresponding value of $\alpha_s$ evaluated at the scale $\mu$. 

\b To be conservative, we have enlarged the original errors of some parameters (gluon condensate, SU(3) breaking,...) by a factor 2-3. We do not consider the lattice result\, \cite{MCNEILE} for $\kappa$ which is in conflict with SR one  from various channels and ChPT. 

We shall see that the effects of the gluon and four-quark condensates on the values of the decay constants are small though they play an crucial r\^ole in the stability analysis. 

\section{Parametrization of the spectral function}
 \b {\it Minimal Duality Ansatz (MDA)} \\
 We shall  use MDA for parametrizing the spectral function:
\bea
\hspace*{-0.7cm}\frac{1}{\pi}\mbox{ Im}\psi_{P/S}(t)&\simeq& f^2_PM_{P/S}^4\delta(t-M^2_P)
  + 
  ``\mbox{QCD cont.}" \theta (t-t^P_c)~,\nnb\\
\hspace*{-0.7cm}  \frac{1}{\pi}\mbox{ Im}\Pi^T_V(t)&\simeq& f^2_VM_V^2\delta(t-M^2_V)
  + 
  ``\mbox{QCD cont.}" \theta (t-t^V_c),
\label{eq:duality}
\eea
where $f_{P,V}$ are the decay constants defined in Eq. (\ref{eq:fp}) and the higher states contributions are smeared by the ``QCD continuum" coming from the discontinuity of the QCD diagrams and starting from a constant threshold $t^P_c,~t^V_c$ which is independent on the subtraction point $\mu$ in this standard minimal model. 

{\b \it Test of the Minimal Duality Ansatz from $J/\psi$ and $\Upsilon$}\label{sec:duality}\\
\nin
The MDA presented in Eq. (\ref{eq:duality}), when applied to the $\rho$-meson reproduces within 15\% accuracy the ratio ${\cal R}_{\bar dd}$ measured from the total cross-section $e^+e^-\to {\rm I=1 ~hadrons}$ data (Fig. 5.6 of \cite{SNB2}), while in the case of charmonium, $M_\psi^2$ from ratio of moments  ${\cal R}^{(n)}_{\bar cc}$ evaluated at $Q^2=0$ has been compared with the one from complete data where a remarkable agreement for higher $n\geq 4$ values (Fig. 9.1 of \cite{SNB2}) has been found. 
 
\begin{figure}[hbt] 
\begin{center}
{\includegraphics[width=4.2cm  ]{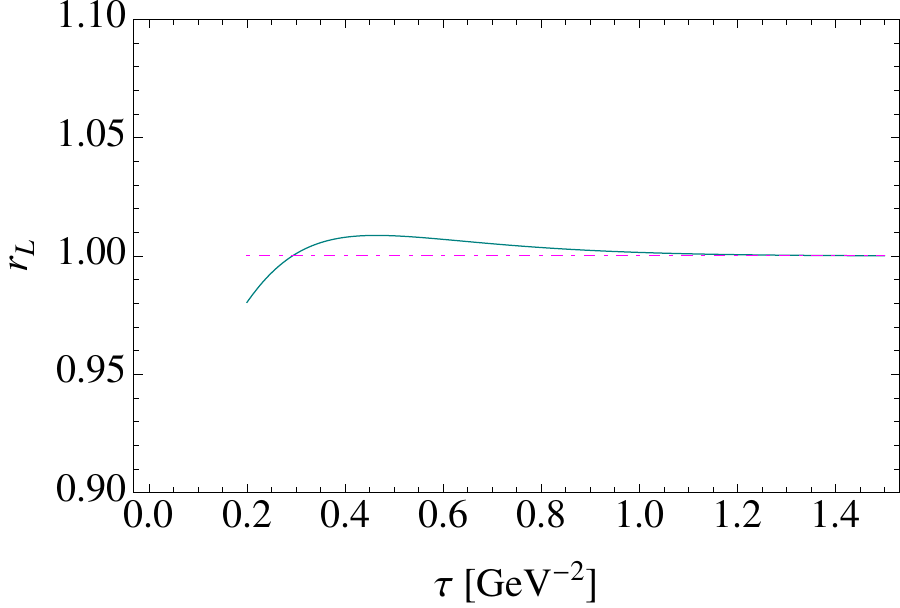}}
{\includegraphics[width=4.2cm  ]{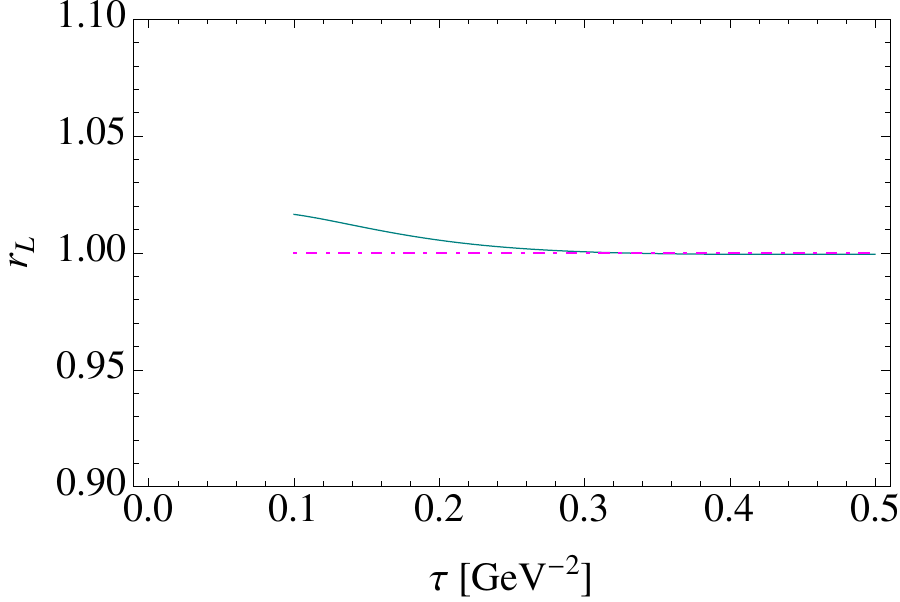}}
\centerline {\hspace*{0cm} \scriptsize a) \hspace*{4cm} b) }\vspace{-0.3cm}
\caption{
\scriptsize 
 $\tau$-behaviour of the ratio of  ${\cal L}^{exp}_{\bar bb}/ {\cal L}^{dual}_{\bar bb}$ for $\sqrt{t_c}= M_{\Upsilon(2S)}$-0.15 GeV. The red dashed curve corresponds to the strict equality for all values of $\tau$ :  a) charmonium,  b) bottomium channels.
}
\label{fig:bduality} 
\end{center}
\end{figure}

Recent tests of MDA from the $J/\psi$ and $\Upsilon$ systems have been done in \cite{SNFB12a}. Taking 
$({t^\psi_c})^{1/2}\simeq M_{\psi(2S)}- 0.15$ GeV and $({t^\Upsilon_c})^{1/2}\simeq M_{\Upsilon(2S)}- 0.15$ GeV, we show (for instance)  the ratio between ${\cal L}^{exp}_{\bar QQ}$ and  ${\cal L}_{\bar QQ}^{dual}$ in Fig.\,\ref{fig:bduality} for the $J/\pi$ and $\Upsilon$ systems indicating that for heavy quark systems the r\^ole of the QCD continuum is smaller than in the case of light quarks while the exponential weight suppresses efficiently the QCD continuum contribution and enhances the one of the lowest resonance.
We have used the simplest QCD continuum  for massless quarks 
\cite{SNH10}\,\footnote{We have checked that the 
including 
mass corrections give the same results.}:
$$
 {\rm QCD~ cont.}=(1+a_s+1.5a_s^2-12.07a_s^3)\theta(t-t_c).
$$
 One can see in Fig.\,\ref{fig:bduality} that the MDA, with a value of $\sqrt{t_c}$ around the one of the 1st radial excitation mass, describes well the complete data in the region of $\tau$-stability  
of the  sum rules\,\cite{SNH10}:
\beq
  \tau^\psi\simeq (0.8\sim 1.4)~{\rm GeV^{-2}},~~
  ~ \tau^\Upsilon\simeq (0.2\sim 0.4)~{\rm GeV^{-2}},
  \label{eq:tau}
 \eeq
  as we shall see later on. Though it is difficult to estimate with precision the systematic error related to this simple model, this feature indicates the ability of the model for reproducing accurately the data. We expect that the same feature is reproduced for the open-charm and beauty vector meson systems where complete data are still lacking.
 Moreover, MDA  has been also used in \cite{PERIS} 
  in the context of large $N_c$ QCD, where 
 it provides a very good approximation to the observables one computes. 
\nin
\vspace*{-0.7cm}
\begin{figure}[hbt] 
\begin{center}
{\includegraphics[width=4.2cm  ]{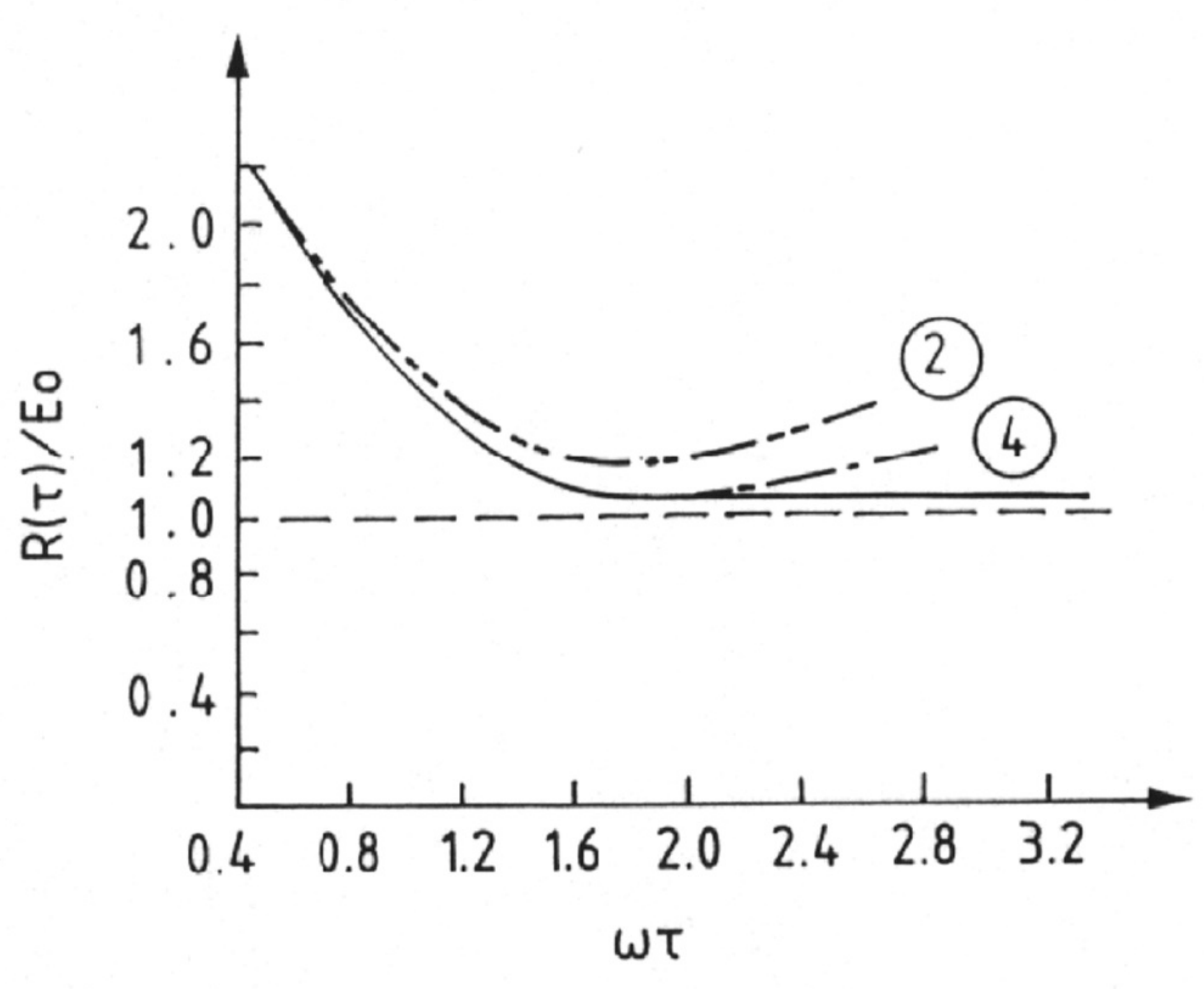}}
{\includegraphics[width=4.2cm  ]{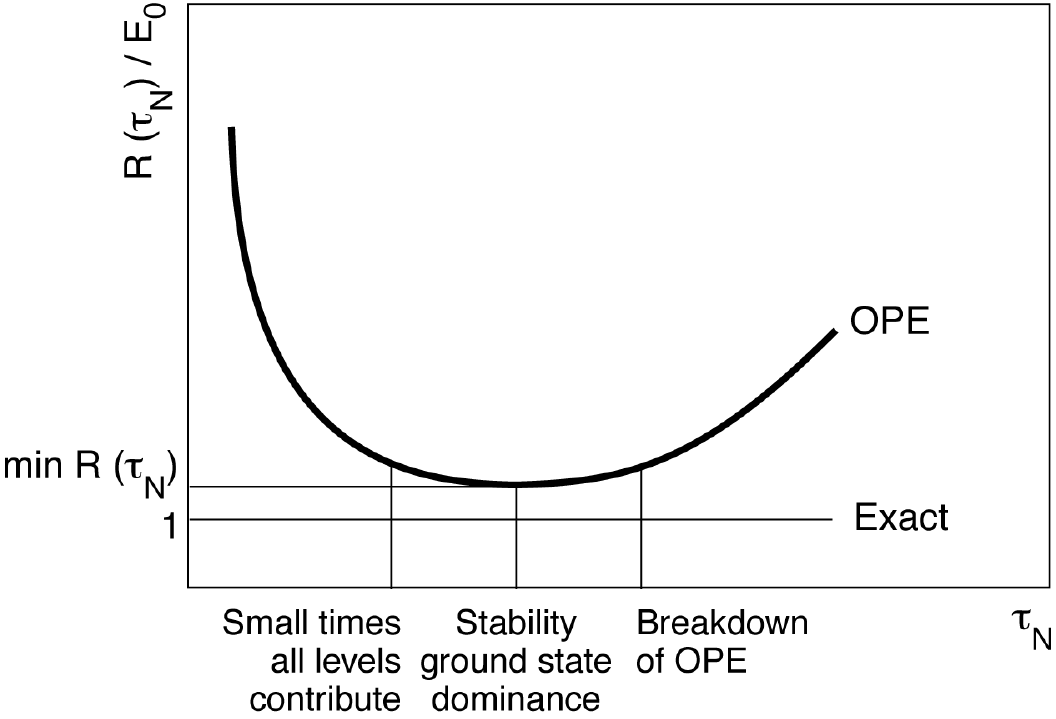}}
\centerline  {\hspace*{0cm} \scriptsize a) \hspace*{4cm} b) }\vspace{-0.3cm}
\caption{
\scriptsize 
{\bf a)} $\tau$-behaviour of ${\cal R}(\tau)$ normalized to the ground state energy $E_0$ for the harmonic oscillator. 2 and 4 indicate the number of terms in the approximate series; {\bf b)} Commented regions for different values of $\tau$.}
\label{fig:oscillo} 
\end{center}
\end{figure} 
\vspace*{-0.7cm}

 \section{Optimization and Stability criteria}
{\b \it $\tau$ and $t_c$-stabilities}\\
In order to extract an optimal information for the lowest resonance parameters from this rather 
crude  description of the spectral function and from the approximate QCD expression, one often applies the stability criteria at which an optimal result can be extracted. This stability is signaled by the existence of a stability plateau, an extremum or an inflexion point versus the changes of the external sum rule variables $\tau$ and $t_c$ where the simultaneous  requirement on the resonance dominance over the continuum contribution and on the convergence of the OPE is satisfied. This optimization criterion demonstrated in series of papers by Bell-Bertmann \cite{BELL} in the case of $\tau$ by taking the examples of harmonic oscillator and charmonium sum rules and extended to the case of $t_c$ in \cite{SNB1,SNB2} gives a more precise meaning of  the so-called ``sum rule window" originally introduced by SVZ \cite{SVZ} and used in the sum rules literature. 

{\b \it $\mu$ subtraction point stability}\\
We shall add to the previous well-known stability criteria, the one associated  to the requirement of stability  versus the variation of the arbitrary subtraction constant $\mu$ often put by hand  in the current literature  and which is often the source of large errors from the PT series in the sum rule analysis.  
The $\mu$-stability procedure has been applied recently in\,\cite{SNFB12a,SNFB12b,SNLIGHT,SNREV14}\,\footnote{Some alternative approaches for optimizing the PT series are in \cite{STEVENSON}.} which gives a much better meaning on the choice of $\mu$-value at which the observable is extracted, while the errors on the results have been reduced due to a better control of the $\mu$ region of variation which is not often the case in the literature.
\section{The decay constants  $f_D$ and $f_{D^*}$}

{\it\b Direct determinations from ${\cal L}_{P,V}(\tau,\mu)$}\\
\begin{figure}[hbt] 
\begin{center}
{\includegraphics[width=4.3cm]{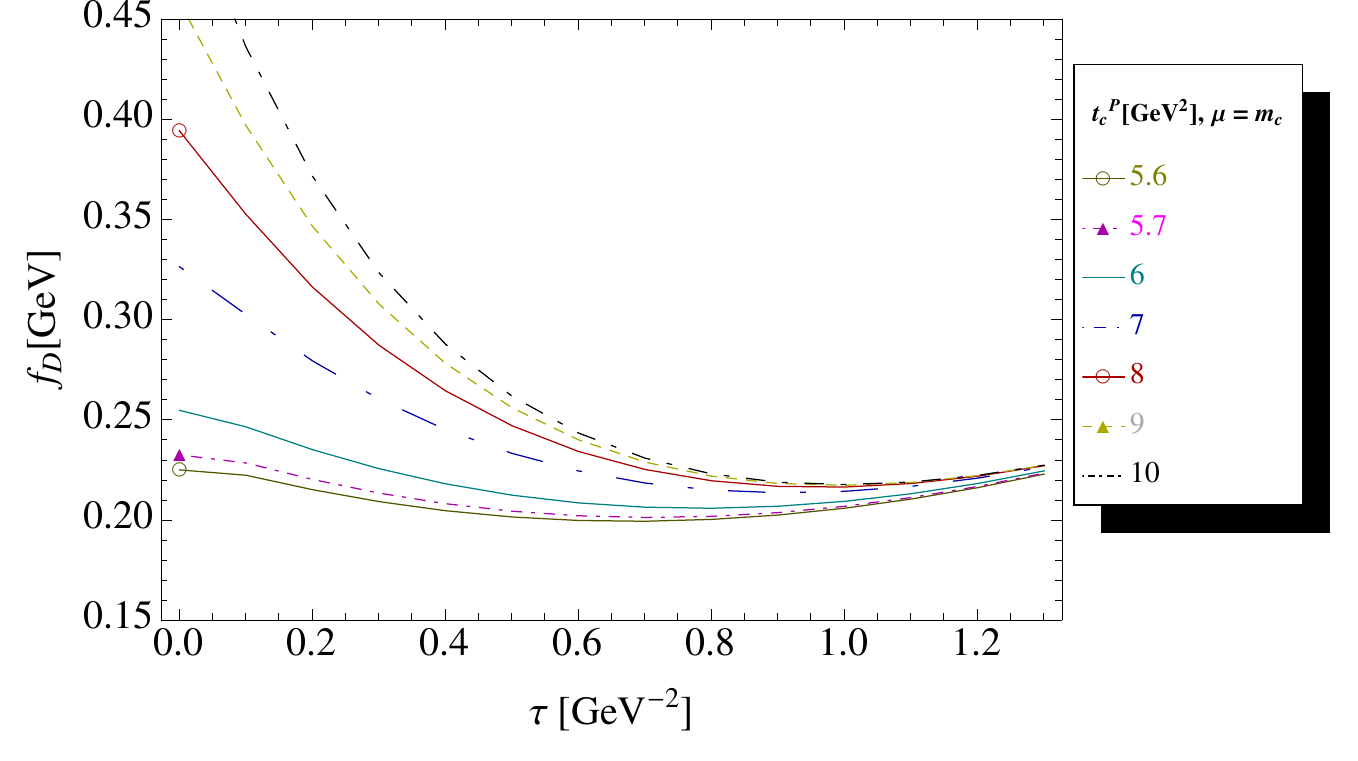}}
{\includegraphics[width=4.3cm]{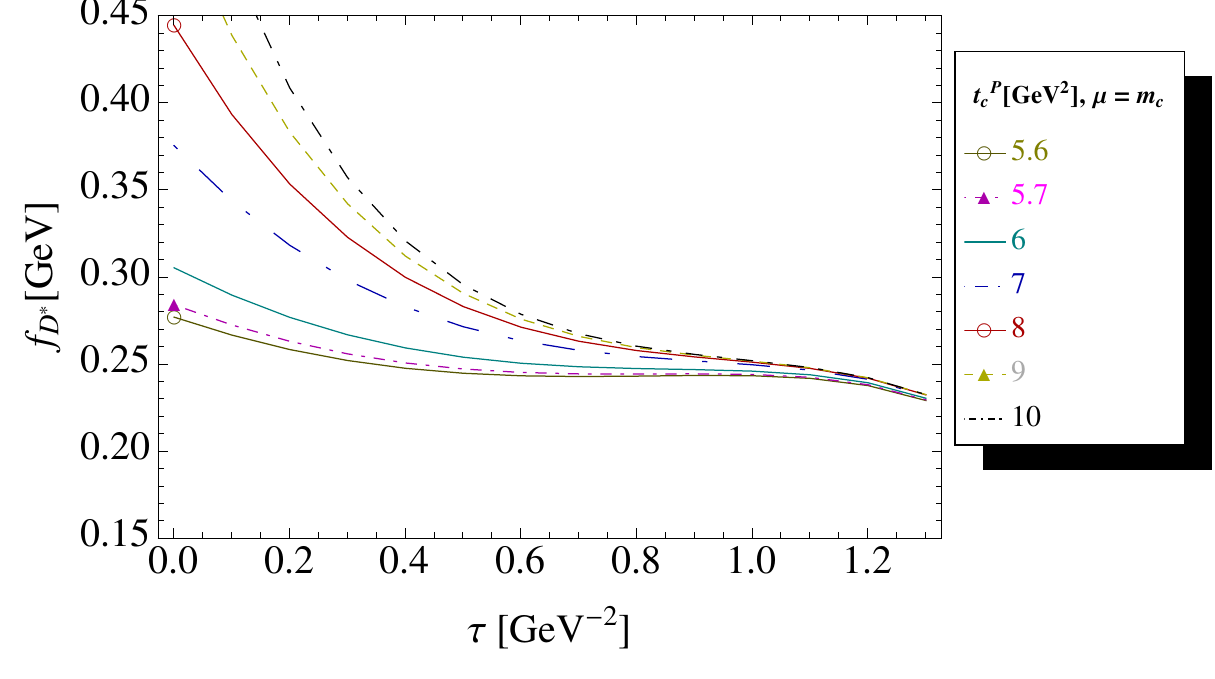}}
\vspace{-0.5cm}
{\hspace*{0.5cm} a) \hspace*{4.5cm} b) }
\end{center}
\caption{
\scriptsize 
{\bf a)} $\tau$-behaviour of $f_{D}$  from  ${\cal L}_{P} $ for different values of $t_c$, at a given value of the subtraction point $\mu=m_c$; {\bf b)} the same as in a) but for $f_{D^*}$ from ${\cal L}_{V} $.}
\label{fig:fdtau} 
\end{figure} 
\nin
We start by showing in Fig. \ref{fig:fdtau}, the $\tau$-behaviour of the decay constants $f_{D^*}$ and $f_D$ at a given value of the subtraction point $\mu=m_c$ for different values of the continuum threshold $t_c$. We have assumed that :
$
({t_c^{D^*}})^{1/2}-({t_c^D})^{1/2}\simeq M_{D^*}-M_{D}=140.6~{\rm MeV},
$
 for the vector and pseudoscalar channels where the QCD expressions are truncated at the same order of PT and NP series. At this value of $\mu$, we deduce
 the optimal value versus $\tau$ (minimum or inflexion point for $\tau\simeq 0.8-1$ GeV$^{-2}$) and $t_c$ (5.6-8 GeV$^2$). Next, we study the $\mu$ variation of these results which we show in Fig. \ref{fig:fd_mean}.
 We deduce the {\it mean result} from \cite{SNFB12a}:
 \beq 
 f_D=204(6)~{\rm MeV}~,
 \label{eq:fd}
  \eeq
 where the largest error for each data point comes from $t_c$\,\footnote{One should note that the extraction of the charm quark running mass by requiring that the sum rules should reproduce the $D$-meson mass is obtained in the same range of values of the previous set of stability parameters \cite{SNFB12a}. The same results are obtained for the other mesons. This fact  increases the confidence on the predictions of the unknown decay constants. }.
The value of $f_{D^*}$  at the minimum in $\mu=(1.5\pm 0.1)$ GeV is \cite{SNFB15}:
 \beq
\hspace*{-0.5cm}f_{D^*}=253.5(11.5)_{t_c}(5.7)_{\tau}(13)_{svz}(1)_\mu=253.5(18.3)~{\rm MeV}, 
\label{eq:fd*1}
\eeq
\bea
{\rm with}&:&
(13)_{svz}=(0.5)_{\alpha_s}(12.3)_{\alpha_s^3}(0.6)_{m_c}(3.8)_{\la\bar dd\ra}\nnb\\
&&(1.8)_{\la \alpha_sG^2\ra}(1.4)_{\la\bar dGd\ra}(0.4)_{\la g^3G^3\ra}(0.4)_{\la\bar dd\ra^2}~,
\eea
where the SVZ-OPE error is dominated by the estimate of the $\alpha_s^3$ corrections (95\%) and where again the error due to $\mu$ has been mutiplied by a factor 2 for a more conservative error.
 One should notice that the accurate value of $f_D$ comes from the mean of different data shown in Fig.\,\ref{fig:fd_mean}, while the error from $f_{D^*}$ is taken from the one at the minimum for $\mu\simeq 1.45$ GeV. 
\begin{figure}[hbt] 
\begin{center}
{\includegraphics[width=4.cm  ]{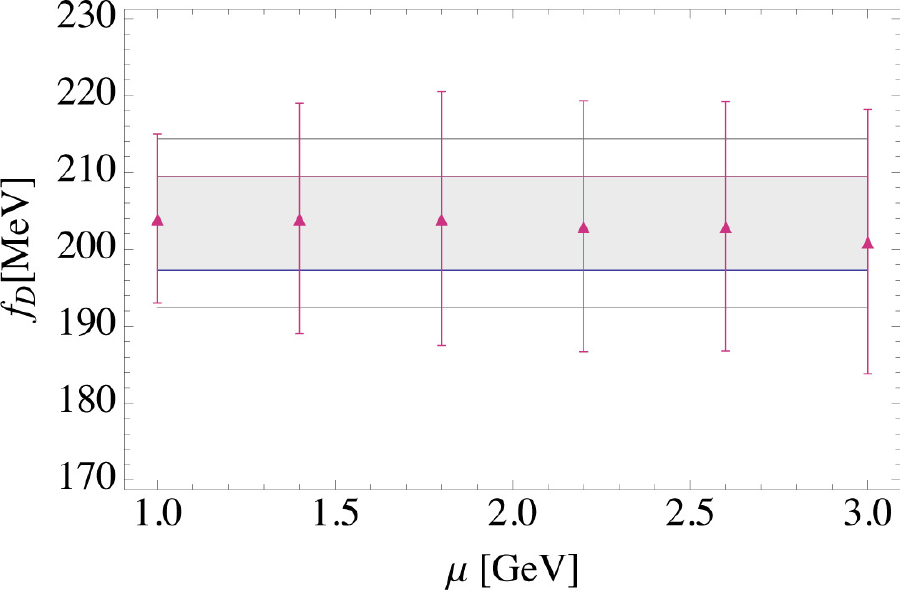}}
{\includegraphics[width=4.6cm  ]{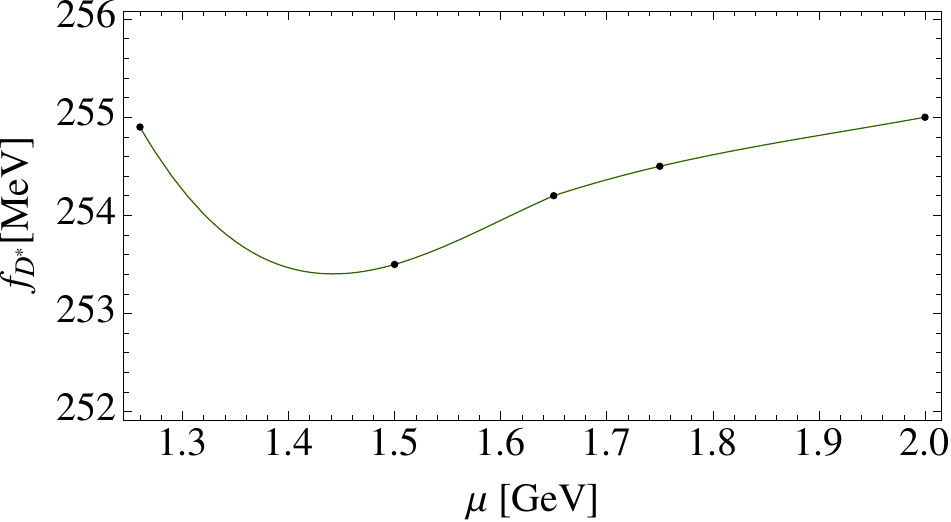}}
\end{center}
\vspace{-0.3cm}
 {\hspace*{0.5cm} a) }\hspace{4cm} b)
\caption{
\scriptsize 
{\bf a)} $f_D$  from LSR versus $\mu$ and for $\hat m_c$=1467 MeV. The filled (grey) region is the average with the corresponding averaged errors. The
dashed horizontal lines are the ones where the errors come from the best determination; {\bf b)} $f_{D^*}$ for different values of $\mu$. }
\label{fig:fd_mean} 
\end{figure} 

\begin{figure}[hbt] 
\begin{center}
{\includegraphics[width=4.2cm  ]{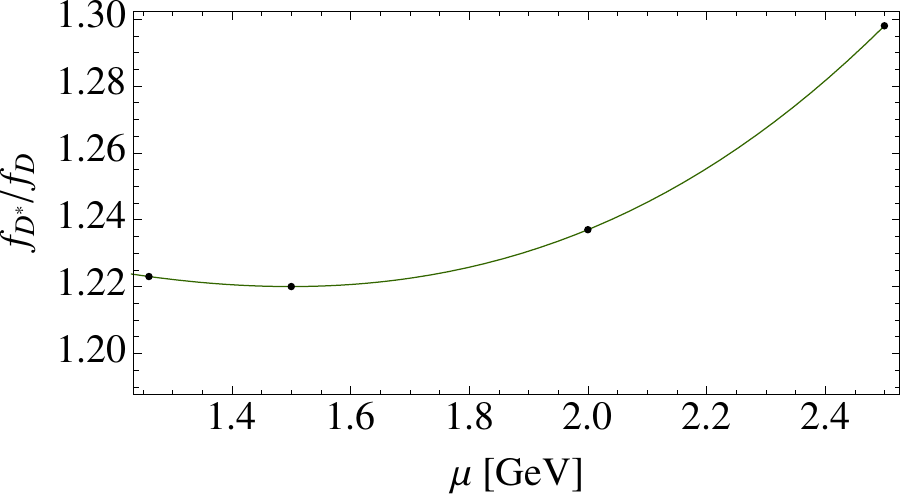}}
{\includegraphics[width=4.2cm  ]{fdst_mu.pdf}}
\end{center}
\vspace{-0.3cm}
 {\hspace*{0.5cm} a) }\hspace{4.cm} b)
\caption{
\scriptsize 
{\bf a)} $f_{D^*}/f_D$  versus the subtraction point $\mu$; {\bf b)} The same as {\bf a)} but for $f_{D^*}$. }
\label{fig:fd*fdmu} \label{fig:fd*mu} 
\end{figure}

{\b \it The ratio $f_{D^*}/f_{D}$ and improved determination of $f_{D^*}$} \\
 One can notice in Fig. \ref{fig:fdtau} that working directly with the ratio in Eq. (\ref{eq:ratio})
by taking the same value $\tau_V=\tau_P$ is inaccurate as the two sum rules ${\cal L}_V(\tau)$ and ${\cal L}_P(\tau)$ are not optimized at the same value of $\tau$ (minimum for $f_D$ and inflexion point for $f_{D^*}$). 
Therefore, for a given value of $t_c$, we take separately the value of each sum rule at the corresponding value of $\tau$
where they present minimum and/or inflexion point and then take their ratio. For a given $\mu$, the optimal result corresponds to the mean obtained in range of values of $t_c$ where one starts to have a $\tau$-stability  ($t_c\simeq 5.6-5.7$ GeV$^2$ for $\tau\simeq 0.6$ GeV$^{-2}$) and a $t_c$-stability ($t_c\simeq 9.5\sim 10.5$ GeV$^2$ for $\tau\simeq 0.8$ GeV$^{-2}$). 
In Fig. \ref{fig:fd*fdmu}, we look for the $\mu$-stability of the previous optimal ratio $f_{D^*}/f_D$ in the set ($\tau,t_c$). We obtain at the minimum $\mu=(1.5\pm 0.1)$ GeV:
\bea
f_{D^*}/f_D&=&1.218(6)_{t_c}(27)_{\tau}(23)_{svz}(4)_\mu~,
=1.218(36) \nnb\\
\label{eq:fdd*}
{\rm with} &:& (23)_{svz}=(7)_{\alpha_s}(2)_{\alpha_s^3}(3)_{m_c}(0)_{\la\bar dd\ra}(18)_{\la \alpha_sG^2\ra}
\nnb\\&&
(12)_{\la\bar dGd\ra}(0)_{\la g^3G^3\ra}(1)_{\la\bar dd\ra^2}~.
\eea
The error from $\mu$ has been multiplied by 2 to be conservative. The largest error comes from $\tau$ which is due to the inaccurate localization of the inflexion point. The one due to $t_c$ is smaller as expected in the ratio which is not the case for the direct extraction of the decay constants. The ones due to ${\la \alpha_sG^2\ra}$ and ${\la\bar dGd\ra}$ are large due to the opposite sign of their contributions in the vector and pseudoscalar channels which add when taking the ratio. 
Using the value $f_D=204(6)$ MeV in Eq. (\ref{eq:fd}) and the ratio in Eq. (\ref{eq:fdd*}), we deduce the improved
value:
\beq
f_{D^*}=248.5(10.4)~{\rm MeV}~,
\label{eq:fd*ratio}
\eeq
where the errors have been added quadratically. 
Taking the mean of the two results in Eqs. (\ref{eq:fd*1}) and (\ref{eq:fd*ratio}), one obtains:
\beq
\la f_{D^*}\ra=249.7(10.5)(1.2)_{syst}
=250(11)~{\rm MeV}~,
\lb{eq:fd*}
\eeq
where the 1st error comes from the most precise determination and the 2nd one from the distance of the mean value to it.

\begin{figure}[hbt] 
\begin{center}
{\includegraphics[width=5cm  ]{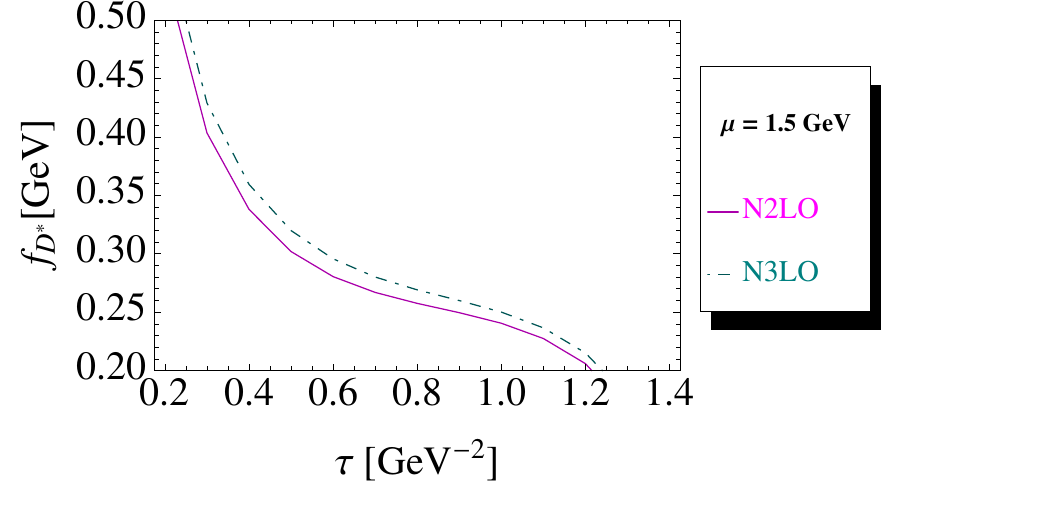}}
{\includegraphics[width=3.75cm  ]{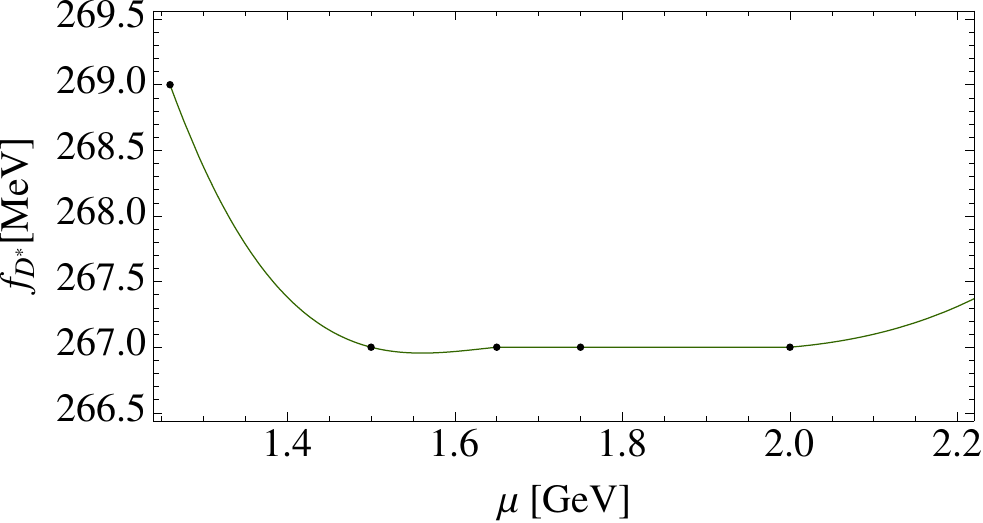}}
\end{center}
\vspace{-0.3cm}
 {\hspace*{0.5cm} a) }\hspace{4.5cm} b)
\caption{
\scriptsize 
Upper bounds on $f_{D^*}$ : {\bf a)}  at different values of $\tau$ for a given value $\mu=$1.5 GeV of the subtraction point $\mu$. One can notice a good convergence of the PT series by comparing the calculated N2LO and estimated N3LO terms; {\bf b)} versus the values of the subtraction point $\mu$.} 
\label{fig:fd*bound_tau} \label{fig:fd*bound_mu} 
\end{figure}

{\it \b Upper bound on $f_{D^*}$} \\
We derive an upper bound on $f_{D^*}$ by considering the positivity of the QCD continuum contribution to the spectral function
and by taking the limit where $t_c\to\infty$ in Eq. (\ref{eq:lsr}) which corresponds to a full saturation of the spectral function by the lowest ground state contribution. The result of the analysis versus the change of $\tau$ for a given value of $\mu=1.5$ GeV is given in Fig.~\ref{fig:fd*bound_tau}a where one can observe like in the previous analysis the presence of a $\tau$-inflexion point. We also show the good convergence of the PT series
by comparing the result at N2LO and the one including an estimate of the N3LO term based on the geometric growth of the PT coefficients. We show in Fig \ref{fig:fd*bound_mu}b the variation of the optimal bound versus the subtraction point $\mu$ where we find a  region of $\mu$ stability from 1.5 to  2 GeV. 
\nin
We obtain:
\bea
f_{D^*}&\leq& 267(10)_\tau(14)_{svz}(0)\mu~{\rm MeV}~,\nnb\\
\label{eq:fdbound1}
{\rm with}&:&
(14)_{svz}=(1.4)_{\alpha_s}(13)_{\alpha_s^3}(1.3)_{m_c}(3)_{\la\bar dd\ra}\nnb\\
&&(3)_{\la \alpha_sG^2\ra}(2.5)_{\la\bar dGd\ra}(0)_{\la g^3G^3\ra}(0.7)_{\la\bar dd\ra^2}~.
\eea
Alternatively, we combine the upper bound $f_D\leq 218.4(1.4)$ MeV obtained in \cite{SNFB12a,SNFB12b} with the ratio in Eq. (\ref{eq:fdd*}) and deduce:
\beq
f_{D^*}\leq 266(8)~{\rm MeV}~,
\label{eq:fdbound2}
\eeq
where we have added the errors quadratically. The good agreement of the results in Eqs. (\ref{eq:fdbound1}) and (\ref{eq:fdbound2}) indicates
the self-consistency of the approaches. This bound is relatively strong compared to the estimate in Eq. (\ref{eq:fd*}).
A comparison of our results with the ones from some other sources is shown in Table\,\ref{tab:res}.
\section{The decay constants $f_B$ and $f_{B^*}$ from SR in full QCD}
\nin
\begin{figure}[hbt] 
\begin{center}
{\includegraphics[width=4.3cm  ]{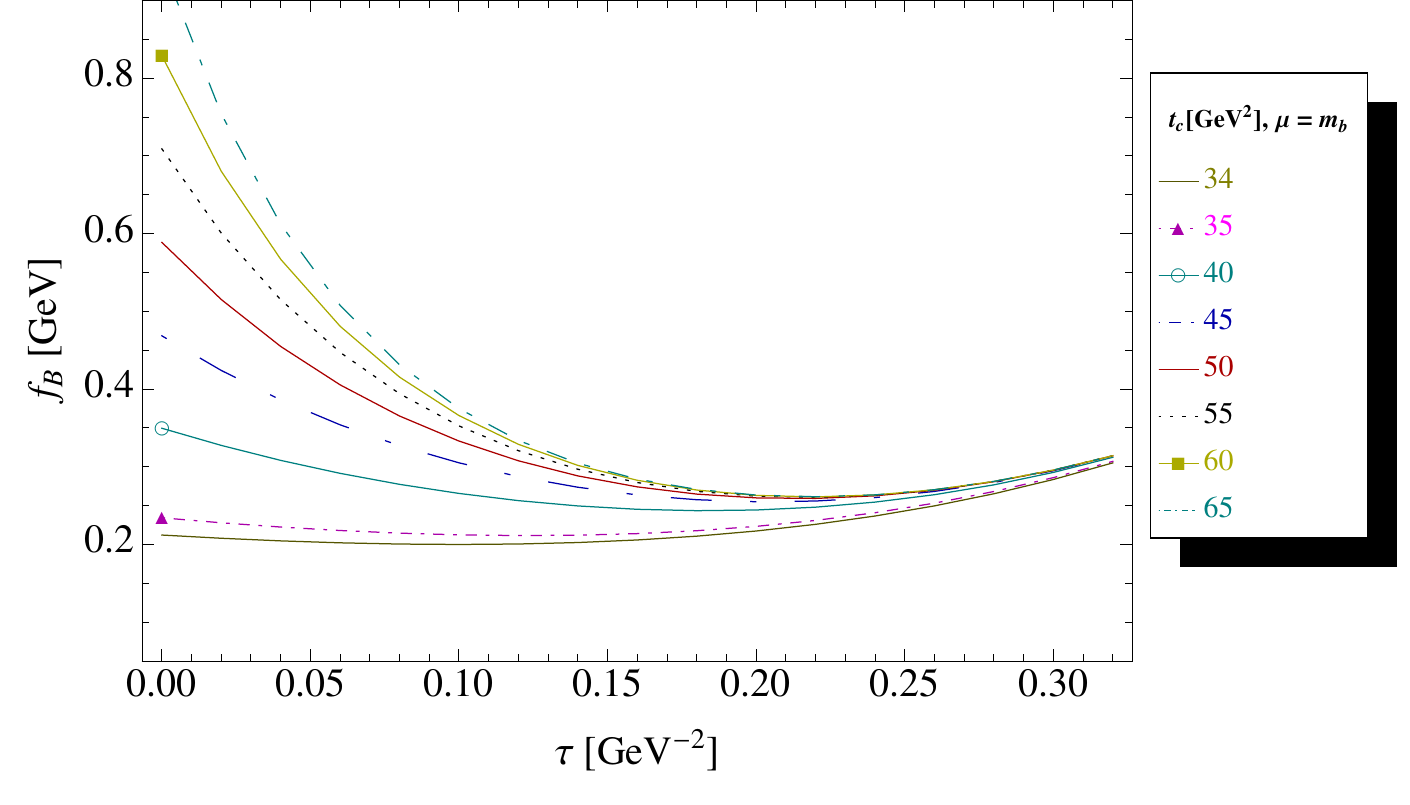}}
{\includegraphics[width=4.3cm  ]{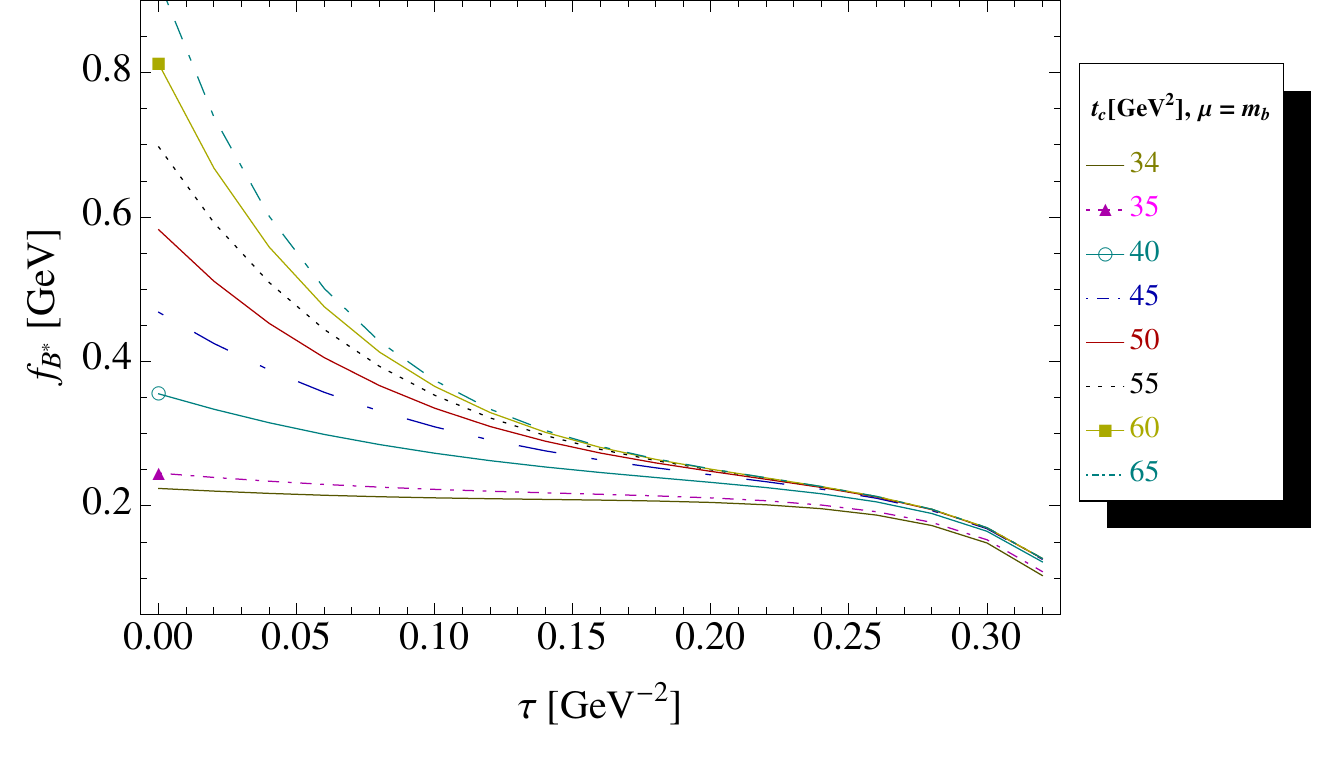}}
\end{center}
\vspace{-0.5cm}
 {\hspace*{0.5cm} a) }\hspace{4.5cm} b)
\caption{
\scriptsize 
{\bf a)} $\tau$-behaviour of $f_{B}$  from  ${\cal L}_{P} $    for different $t_c$, at $\mu=m_b$; {\bf b)} the same as in a) but for $f_{B^*}$ from ${\cal L}_{V} $.}
\label{fig:fbtau} 
\end{figure} 
\nin
{\it\b Direct estimate of  $f_B$ and $f_{B^*}$}\\
We extend the analysis to the case of the $B^*$ meson. We use the set of parameters in Table 1. The $\tau$-behaviours of $f_{B}$ and $f_B^*$ in Fig. \ref{fig:fbtau} have a shape similar to the case of $f_{D}$ and $f_{D^*}$. 
We estimate  $f_{B^*}$ from Fig. \ref{fig:fbtau}b where the $\tau$-stability is reached from $t_c=34$ GeV$^2$ while the $t_c$-one starts from $t_c=(55-60)$ GeV$^2$. We show the $\mu$-behaviour of the optimal
result in Fig. \ref{fig:fb*mu}.  At  the inflexion point $\mu=(5.5\pm 0.5)$ GeV, we deduce:
\bea
f_{B^*}&=&239(38)_{t_c}(1)_{\tau}(2.7)_{svz}(1.4)_\mu
=239(38)~{\rm MeV}, \nnb\\
\label{eq:fb*mu}
{\rm with}&:&
(2.7)_{svz}=(0.7)_{\alpha_s}(2)_{\alpha_s^3}(0.4)_{m_b}(1.6)_{\la\bar dd\ra}\nnb\\
&&(0.4)_{\la \alpha_sG^2\ra}(0.3)_{\la\bar dGd\ra}(0)_{\la g^3G^3\ra}(0)_{\la\bar dd\ra^2}~.
\eea
{\b \it The ratio $f_{B^*}/f_{B}$}\\
We extract  directly the ratio $f_{B^*}/f_B$ from the ratio of sum rules. We show in Fig. \ref{fig:rbst_tau}a its $\tau$-behaviour for different values of $t_c$ for $\mu=5.5$ GeV from which we deduce as optimal value the mean of the $\tau$-minima obtained  from $t_c=34$ to 60 GeV$^2$. We show  in Fig. \ref{fig:rbst_mu}b the $\mu$-behaviour of these optimal results where we find a minimum in $\mu$ around 3.8-4.5 GeV and a slight inflexion point around 5.5 GeV. We consider as a final result the mean of the ones from these two regions of $\mu$:
\beq
{f_{B^*}/ f_{B}}=1.016(12)_{t_c}(1)_\tau(9)_{svz}(6)_\mu
=1.016(16)~,
\lb{eq:rb*mu}
\eeq
with:
\beq
\hspace*{-0.7cm}(9)_{svz}=(3)_{\alpha_s}(6)_{\alpha_s^3}(3)_{m_b}(4)_{\la\bar dd\ra}(3)_{\la \alpha_sG^2\ra}
(1)_{\la\bar dGd\ra}(1)_{\la g^3G^3\ra}(1)_{\la\bar dd\ra^2}
\eeq
Combining the results in Eq. (\ref{eq:rb*mu}) with the value $f_B=206(7)$ MeV obtained in \cite{SNFB12a,SNFB12b}, we deduce:
\beq
f_{B^*}=209(8)~{\rm MeV}~,
\lb{eq:fb*}
\eeq
which is more accurate than the direct determination in Eq. (\ref{eq:fb*mu}) and where the main error comes from the one of $f_B$ extracted in \cite{SNFB12a,SNFB12b}. We consider the result in Eq. (\ref{eq:fb*}) which is also the mean of the results in Eqs. (\ref{eq:fb*}) and (\ref{eq:fb*mu}) as our final determination. 
\begin{figure}[hbt] 
\begin{center}
{\includegraphics[width=4.cm  ]{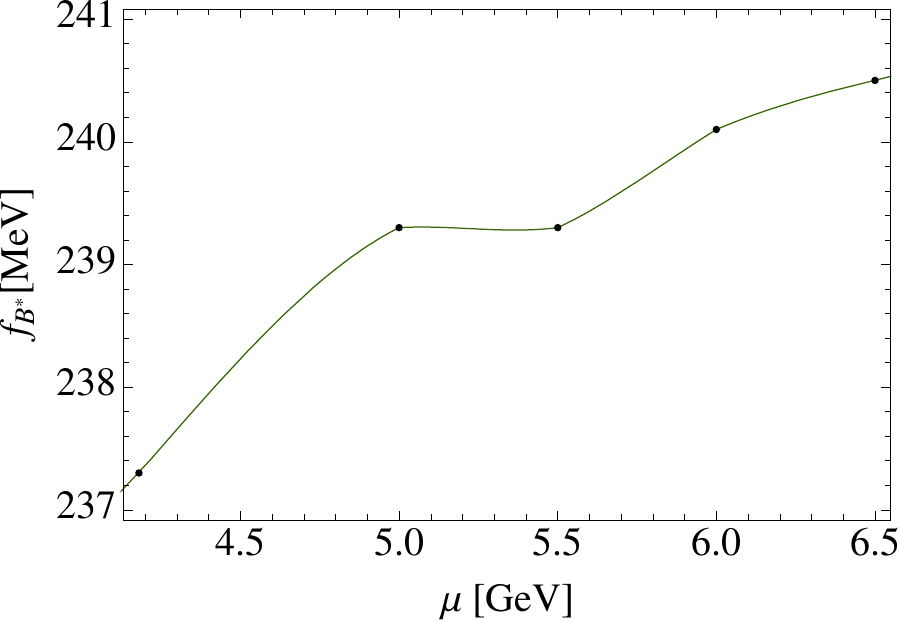}}
\caption{
\scriptsize 
Values of  $f_{B^*}$  at different values of the subtraction point $\mu$. }
\label{fig:fb*mu} 
\end{center}
\end{figure} 
\vspace*{-0.5cm}
\begin{figure}[hbt] 
\begin{center}
{\includegraphics[width=4.3cm  ]{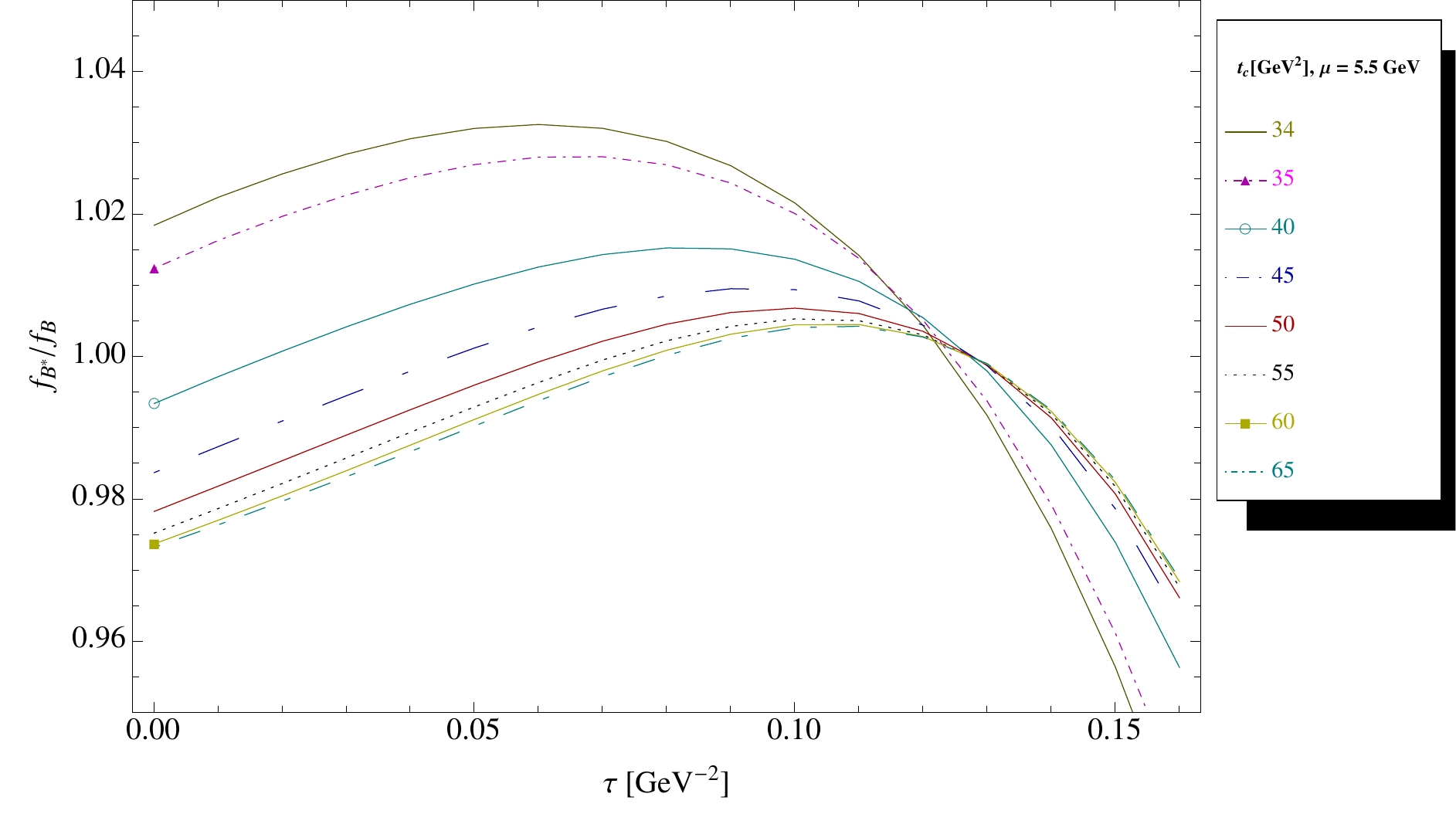}}
{\includegraphics[width=4.3cm  ]{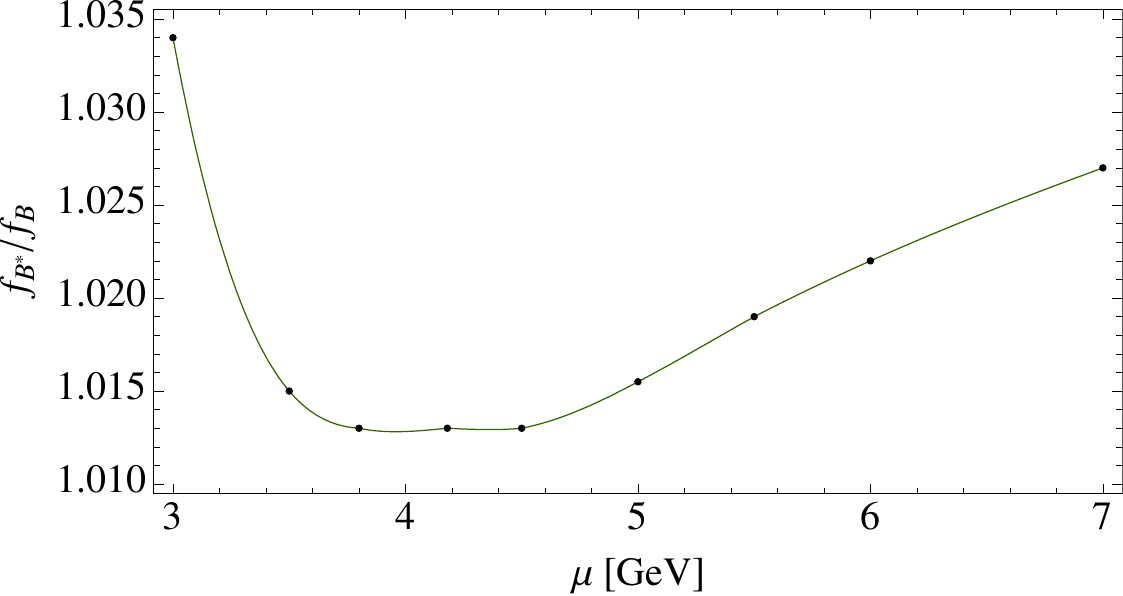}}
\end{center}
\vspace{-0.5cm}
 {\hspace*{0.5cm} a) }\hspace{4.2cm} b)
\caption{
\scriptsize 
$f_{B^*}/f_B$: {\bf a)} $\tau$-behaviour for different $t_c$ at $\mu=5.5$ GeV;  {\bf b)}   $\mu$-behaviour. }
\label{fig:rbst_tau} \label{fig:rbst_mu} 
\end{figure} 
{\it\b Upper bound on $f_{B^*}$}\\
Like in the case of the $D^*$ meson, we extract directly an upper bound on $f_{B^*}$ by using the positivity of the QCD continuum to the spectral function. We show the $\tau$- and the $\mu$-behaviours of the optimal bound in Fig. \ref{fig:fbbound_mu}.
We deduce at $\mu=(6\pm 0.5)$ GeV:
\beq
f_{B^*}\leq 295(14)_\tau (4)_{svz}(10)_\mu=
295(18)~{\rm MeV}~.
\label{eq:fb*_bound}
\eeq
\begin{figure}[hbt] 
\begin{center}
{\includegraphics[width=5.2cm  ]{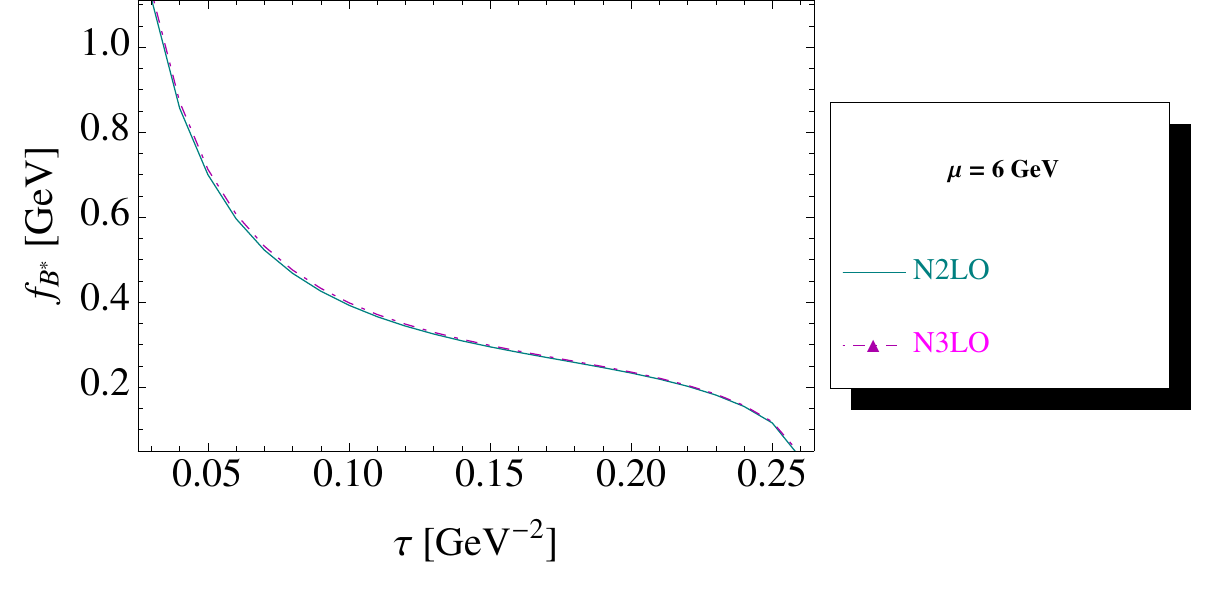}}
{\includegraphics[width=3.5cm  ]{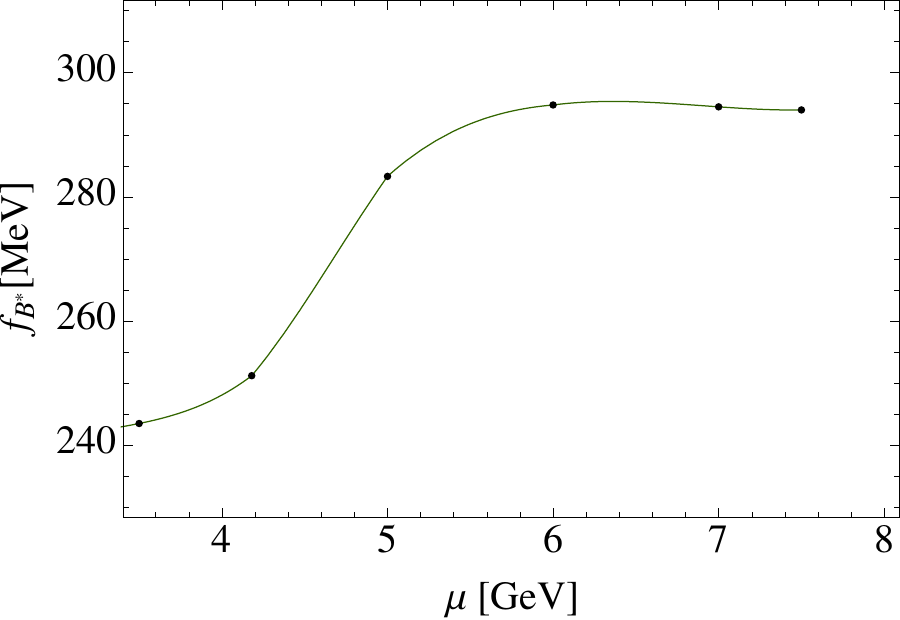}}
\end{center}
\vspace{-0.5cm}
 {\hspace*{0.5cm} a) }\hspace{5.2cm} b)
\caption{
\scriptsize 
Upper bound of $f_{B^*}$  from  ${\cal L}_{P}$:  {\bf a)} $\tau$-behaviour at $\mu=6$ GeV for N2LO and N3LO truncation of the PT series; {\bf b)}  $\mu$-behaviour.}
\label{fig:fbbound_tau} \label{fig:fbbound_mu}
\end{figure} 
\nin
We consider the previous values of $f_{D^*}$ and $f_{B^*}$ as improvement of our earlier results in \cite{SNB2} and \cite{SNFB88}. 
\section{ $f_{B^*}/f_{B}$ from HQET sum rules}
This ratio is under a good control from heavy quark effective theory (HQET) sum rules. It reads in the large mass limit  \cite{BEC,BALL,NEUBERT2}:
\bea
R_b&\equiv&{f_{B^*}\sqrt{M_{B^*}}\over f_{B}\sqrt{M_{B}}}= R_{PT}+{R_{NPT}\over M_b}~,\nnb\\
 R_{PT}&=&1-{2\over 3}a_s -6.56a_s^2-84.12a_s^3 = 0.892(13)~,\nnb\\
 R_{NPT}&=& {2\over 3}\bar\Lambda\ga 1+{5\over 3}a_s\dr- 4G_{\Sigma}\ga 1+{3\over 2}a_s\dr~,
 \eea
 with : $a_s\equiv (\alpha_s/\pi)(M_b)=0.070(1)$, $\bar\Lambda \equiv  M_B-M_b=(499\pm 60)$~MeV, where we have used the pole mass $M_b$=4.810(60)~MeV  to order $\alpha_s^2$ deduced from the running mass $\bar m_b(m_b)=4.177(11)$ MeV \cite{SNH10} and we have retained the larger error from PDG \cite{PDG}. We have estimated the error in the PT series by assuming a geometric growth of the coefficients \cite{NZ}. 
\beq 
G_{\Sigma}(M_b)\equiv \hat G_\Sigma [\alpha_s(M_b)]^{-3/2\beta_1}+{8\over 27}\bar\Lambda~,
\eeq
where the invariant quantity: $\hat G_\Sigma  \simeq - (0.20\pm 0.01)$ GeV has been extracted inside   the $t$-stability region of Fig. 7  from \cite{BALL}.  We obtain :
\beq
R_b=0.938(13)_{PT}(5)_{NP}(6)_{M_b,\bar\Lambda}=0.938(15)~.
\eeq
To convert the above HQET result to the one of the full theory at finite quark and meson masses, we have to include the normalization factor $(M_B/M_b)^2$ relating the pseudoscalar to the universal HQET correlators  according to the definition in Eq.\,(\ref{eq:fp}).  Then, we deduce (see e.g. \cite{SNFB12b,BALL,PENIN}): 
\beq
{f_{B^*}\over f_B}= \ga{M_B\over {M_b}}\dr \sqrt{M_B\over M_{B^*}}R_b=1.025(13)_{M_b}(15)_{R_b}~,
\label{eq:ratiohqet}
 \eeq 
 in fair agreement with our result in Eq. (\ref{eq:rb*mu}) and the ones in \cite{PIVOV,BECIR2,BALL} but higher than the ones in \cite{LUCHA2,DAVIES}. To make a direct comparison of our results with the lattice calculations, it is desirable to have a direct lattice calculation from the pseudoscalar correlator built from the current in Eq. (\ref{eq:fp}).  
 The result of \cite{LUCHA2} is difficult to interpret as they use a non-standard threshold of the QCD continuum. However, their requirement of maximal stability for $\tau\leq 0.15 $ GeV$^{-2}$ is outside  the ``standard sum rule (SR) window" obtained around 0.2-0.3 GeV$^{-2}$
(minimum or inflexion point in our figures) where the lowest resonance dominates the sum rule.    
 Taking literally their set $(t_c,\tau)$=( 33 GeV$^2$, 0.05 GeV$^{-2}$) where their ``duality" is obtained (Fig. 3 of \cite{LUCHA2})  into our Fig. \ref{fig:rbst_tau}, one obtains: $f_{B^*}/f_B=.985$ more comparable with their result 0.944 but meaningless from the SR point of view as it comes from a region dominated by the
 QCD continuum. 
\section{ SU(3) breaking for $f_{D^*_s}$ and $f_{D^*_s}/f_{D^*}$ }
\begin{figure}[hbt] 
\begin{center}
{\includegraphics[width=4.3cm  ]{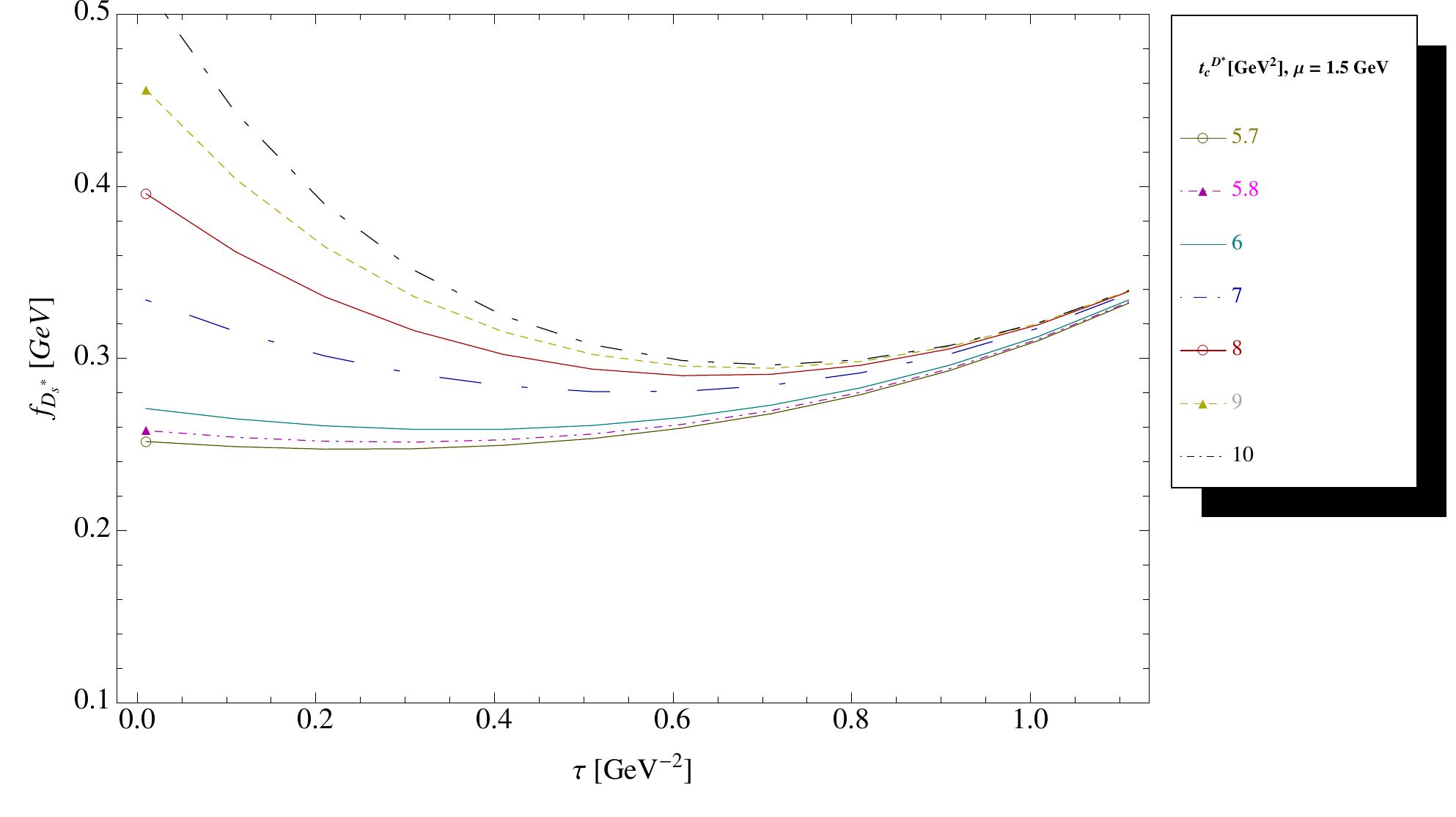}}
{\includegraphics[width=4.3cm  ]{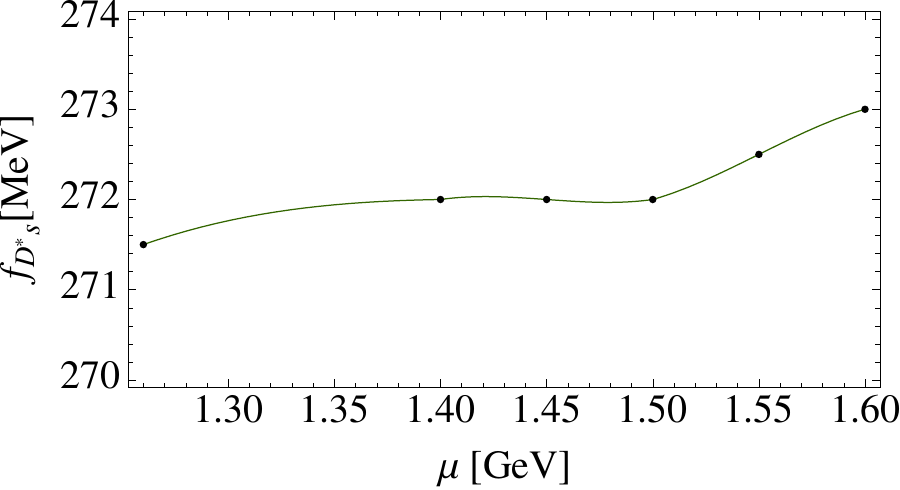}}
\end{center}
\vspace{-0.5cm}
 {\hspace*{0.5cm} a) }\hspace{4.3cm} b)
\caption{
\scriptsize 
$f_{D^*_s}$ from  ${\cal L}_{V}$: {\bf a)}  $\tau$-behaviour  for different $t_c$, at  $\mu=$1.5 GeV; {\bf b)}  $\mu$-behaviour.}
\label{fig:fd*stau} \label{fig:fd*smu} 
\end{figure}

We pursue the same analysis for studying the $SU(3)$ breaking for  $f_{D^*_s}$ and the ratio $f_{D^*_s}/f_{D^*}$. We work with the complete massive $(m_s\not=0)$ LO expression of the PT spectral function obtained in \cite{FNR} and the massless $(m_s=0)$ expression known to N2LO used in the previous sections.  We include the NLO PT corrections due to linear terms in $m_s$ obtained in \cite{PIVOV}.   We show the $\tau$-behaviour of the results in Fig.~\ref{fig:fd*stau}a for a given $\mu=1.5$ GeV and different $t_c$. We study their $\mu$-dependence  in Fig.~\ref{fig:fd*smu}b where a nice $\mu$ stability is reached for $\mu\simeq 1.4-1.5$ GeV. 
\nin
We have used :
$({t_c^{D^*_s}})^{1/2}-({t_c^{D^*}})^{1/2}\approx M_{D^*_s}-M_{D^*}=102~{\rm MeV}.
$
Taking the conservative result ranging from the beginning of $\tau$-stability ($t_c\simeq 5.7$ GeV$^2$) until the beginning of $t_c$-stability of about (9--10) GeV$^2$, we obtain at $\mu$=1.5 GeV:
\bea
f_{D^*_s}&=&272(24)_{t_c}(2)_\tau(18)_{svz}(2)_\mu 
= 272(30)~{\rm MeV}~,\nnb\\
\lb{eq:fd*s}
{\rm with}&:&
(18)_{svz}=(0)_{\alpha_s}(14)_{\alpha_s^3}(1)_{m_c}(2)_{\la\bar dd\ra}(1.5)_{\la \alpha_sG^2\ra}
\nnb\\
&&(0.8)_{\la\bar dGd\ra}(0)_{\la g^3G^3\ra}(0)_{\la\bar dd\ra^2}
(0.3)_{m_s}(2)\kappa~.
\eea
Taking the PT linear term in $m_s$ at lowest order and $t_c=7.4$ GeV$^2$, we obtain $f_{D^*_s}=291$ MeV in agreement with the one 293 MeV of \cite{PIVOV} obtained in this way. The inclusion of the complete LO term 
decreases this result by about 5 MeV while the inclusion of the NLO PT $SU(3)$ breaking terms increases the result by about the same amount. 
Combining the result in Eq. (\ref{eq:fd*s}) with the one in Eq. (\ref{eq:fd*}), we deduce the ratio:
\beq
f_{D^*_s}/ f_{D^*}=1.090(70)~,
\lb{eq:rd*stau1}
\eeq
\begin{figure}[hbt] 
\begin{center}
{\includegraphics[width=4.3cm  ]{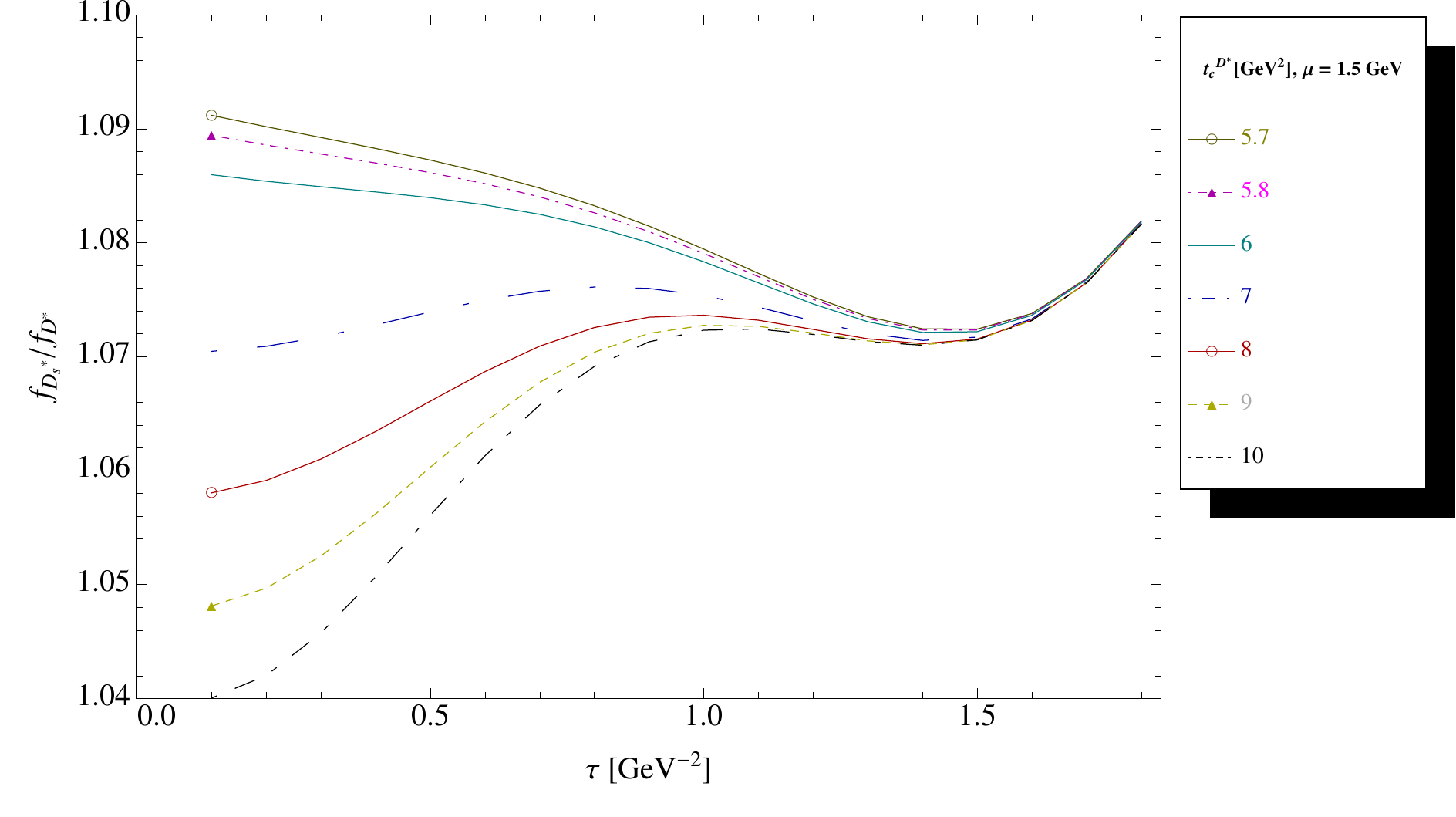}}
{\includegraphics[width=4.3cm  ]{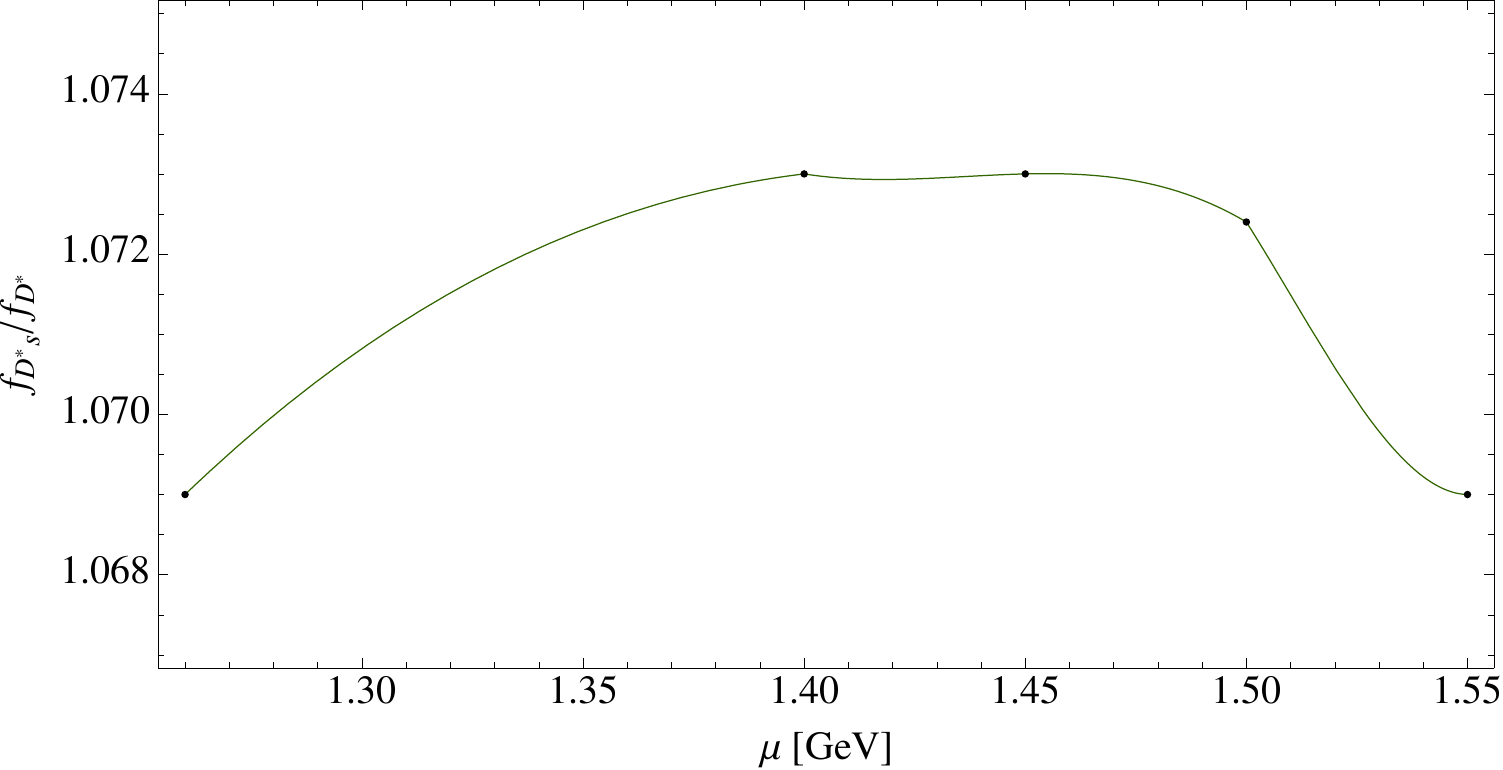}}
\end{center}
\vspace{-0.5cm}
 {\hspace*{0.5cm} a) }\hspace{4.3cm} b)
\caption{
\scriptsize 
$f_{D^*_s}/f_{D^*}$: {\bf a)}  $\tau$-behaviour  for different
$t_c$ and for $\mu=1.5$ GeV; 
{\bf b)}  $\mu$-behaviour.}
\label{fig:rd*stau} \label{fig:rd*smu}
\end{figure} 
\nin
where we have added the relative errors quadratically. Alternatively, we extract directly the previous ratio using the ratio of sum rules. We show the results in Fig. \ref{fig:rd*stau}a versus $\tau$ and for different values of $t_c$ at $\mu=1.5$ GeV. $\tau$-stabilities occur from $\tau\simeq$ 1 to 1.5 GeV$^{-2}$. We also show in Fig \ref{fig:rd*smu}b the $\mu$-behaviour of the results where a good stability in $\mu$ is observed for $\mu \simeq (1.4-1.5)$ GeV in the same way as for $f_{D^*_s}$. We deduce:
\bea
f_{D^*_s}/ f_{D^*}&=&1.073(1)_{t_c}(16)_\tau(2)_\mu(50)_{svz}
=1.073(52)~,\nnb\\
\lb{eq:rd*stau2}
{\rm with}&:&(50)_{svz}=(1)_{\alpha_s}(45)_{\alpha_s^3}(0)_{m_c}(2)_{\la\bar dd\ra}(16)_{\la \alpha_sG^2\ra}\nnb\\
&&
(3)_{\la\bar dGd\ra}(1)_{\la g^3G^3\ra}(2)_{\la\bar dd\ra^2}
(4)_{m_s}(13)\kappa~,
\eea
where, for asymmetric errors, we have taken the mean of the two extremal values. The error associated to $\tau$ take into accounts the fact that, for some values of $t_c$, the $\tau$-minima for $f_{D^*}$ and $f_{D^*_s}$ do not co\"\i ncide. Comparing Eqs. (\ref{eq:rd*stau1}) and  (\ref{eq:rd*stau2}), one can see the advantage of a direct extraction from the ratio of moments due to the cancellation of systematic errors in the analysis. Taking the mean of Eqs. (\ref{eq:rd*stau1}) and  (\ref{eq:rd*stau2}),   we deduce:
\beq
f_{D^*_s}/ f_{D^*}=1.08(6)(1)_{syst}~,
\lb{eq:fd*sd*}
\eeq
where the 1st error comes from the most precise determination and the 2nd one from the distance of the mean value to the central value of this precise determination. 
Using Eq. (\ref{eq:fd*sd*}), $f_{D^*}$ in Eq. (\ref{eq:fd*}) and its upper bound in Eq. (\ref{eq:fdbound2}), we predict in MeV:
\beq
f_{D^*_s}=270(19)~,~~~~f_{D^*_s}\leq 287(8.6)(16)
= 287(18)~.
\eeq
Future experimental measurements of  $f_{D^*}$ and $f_{D^*_s}$ though most probably quite difficult should provide a decisive selection of these existing theoretical predictions.
\vspace*{-0.25cm}
\section{ SU(3) breaking for $f_{B^*_s}$ and $f_{B^*_s}/f_{B^*}$ }
\vspace*{-0.25cm}
\begin{figure}[hbt] 
\begin{center}
{\includegraphics[width=4.3cm  ]{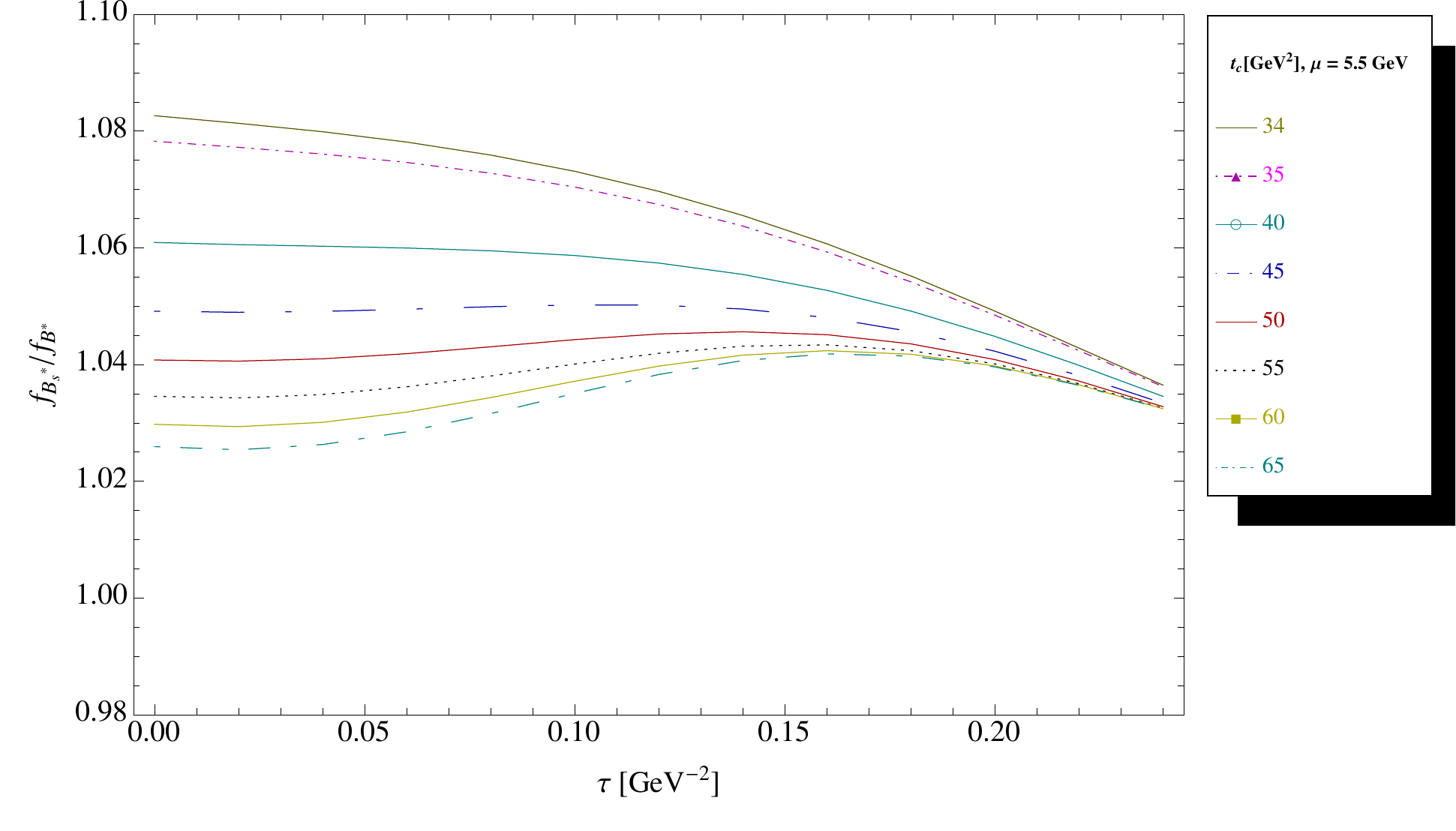}}
{\includegraphics[width=4.3cm  ]{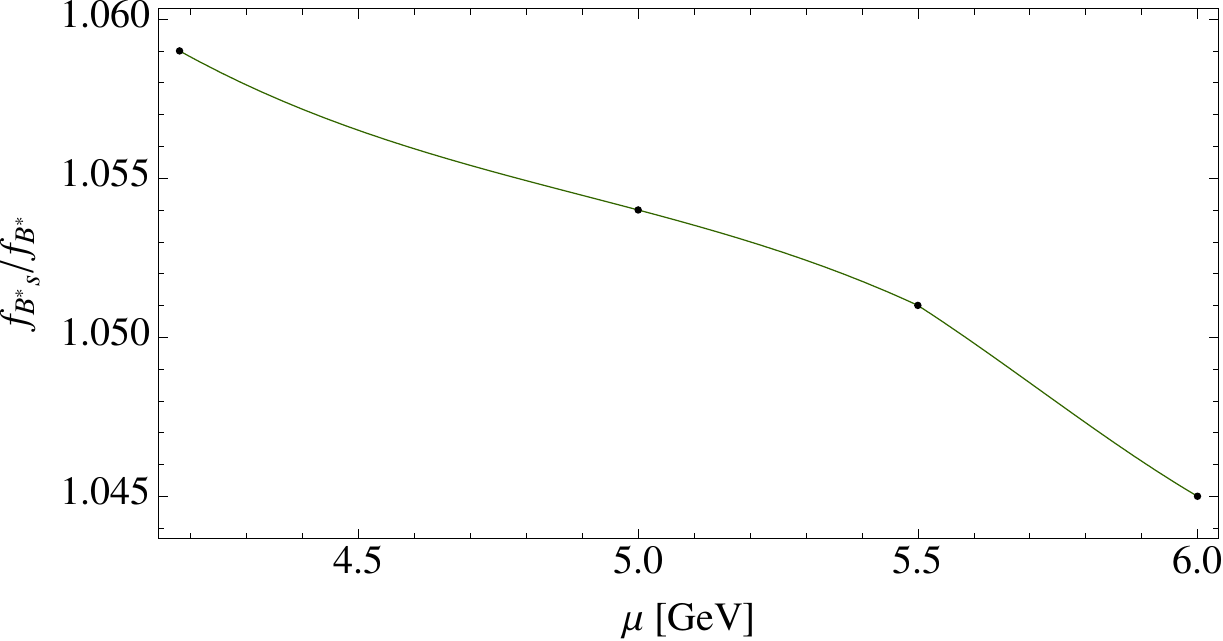}}
\end{center}
\vspace{-0.5cm}
 {\hspace*{0.5cm} a) }\hspace{4.2cm} b)
\caption{
\scriptsize 
$f_{B^*_s}/f_{B^*}$: {\bf a)} $\tau$-and $t_c$-behaviours 
at a given  
$\mu=5$ GeV; {\bf b)} $\mu$-behaviour.}
\label{fig:rb*stau} \label{fig:rb*smu} 
\end{figure} 
\nin
We extend the analysis done for the $D^*_s$ to the case of the $B^*_s$-meson. 
We show, in Fig. \ref{fig:rb*stau}a, the $\tau$-behaviour of the ratio $f_{B^*_s}/f_{B^*}$  at $\mu$=5 GeV and for different values of $t_c$ where the $\tau$-stability starts from $t_c=40$ GeV$^2$
while the $t_c$ one is reached for $t_c\simeq (60-65)$ GeV$^2$. Our optimal result is taken in this range of $t_c$. We study the $\mu$-behaviour in Fig. \ref{fig:rb*smu}b where  an inflexion point is obtained for $\mu=(5\pm.5)$ GeV. At this point, we obtain:
\bea
f_{B^*_s}/f_{B^*}&=&1.054(8)_{t_c}(0)_\tau(6)_\mu(4.6)_{svz}
=1.054(11)~,\nnb\\
\lb{eq:rb*stau}
{\rm with}&:&
(4.6)_{svz}=(2)_{\alpha_s}(2.5)_{\alpha_s^3}(0)_{m_b}(2)_{\la\bar dd\ra}(1.5)_{\la \alpha_sG^2\ra}\nnb\\
&&(0)_{\la\bar dGd\ra}(0)_{\la g^3G^3\ra}(0)_{\la\bar dd\ra^2}(0)_{m_s}(2)\kappa~.
\eea
\begin{figure}[hbt] 
\begin{center}
{\includegraphics[width=4.3cm  ]{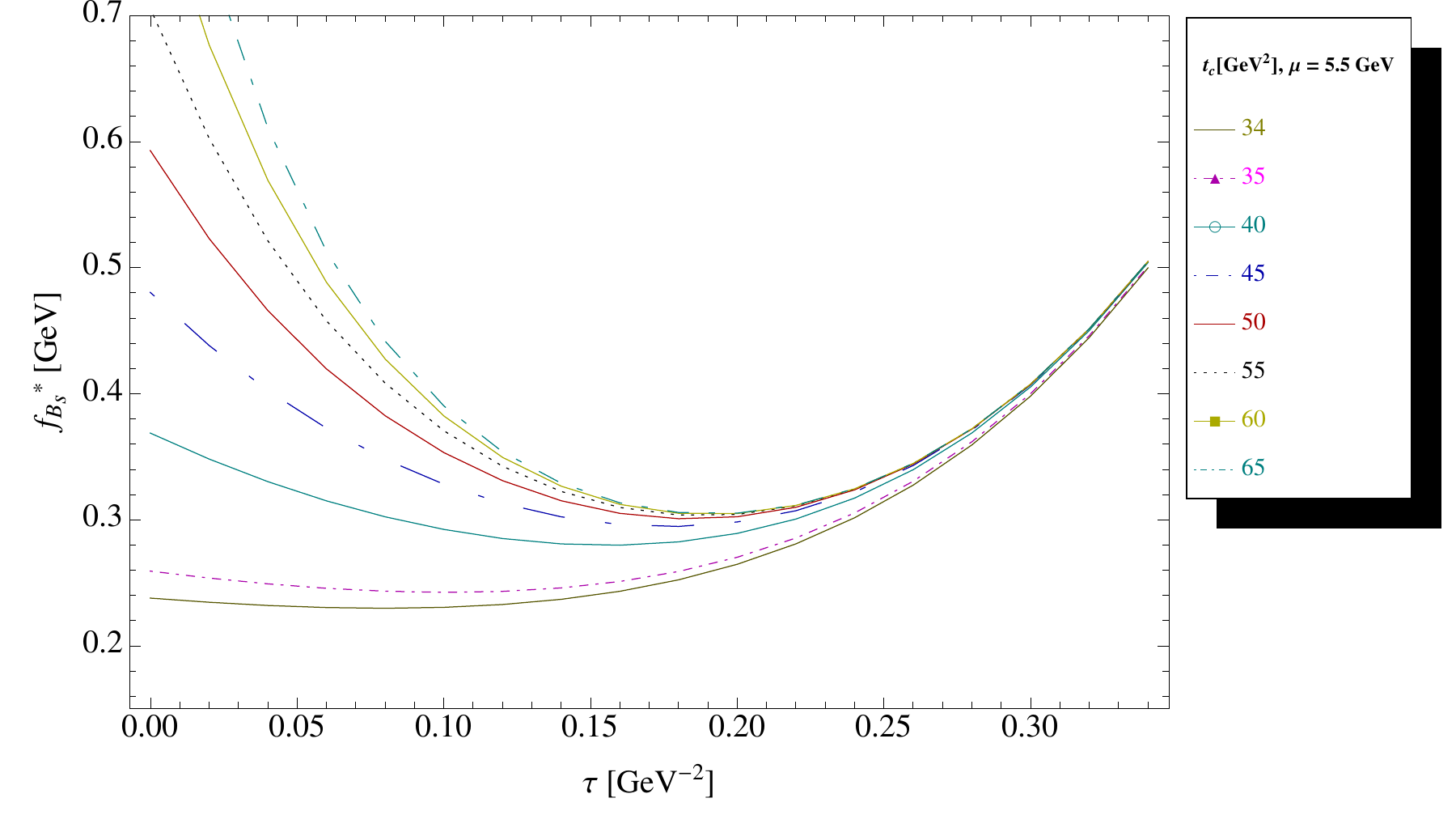}}
{\includegraphics[width=4.3cm  ]{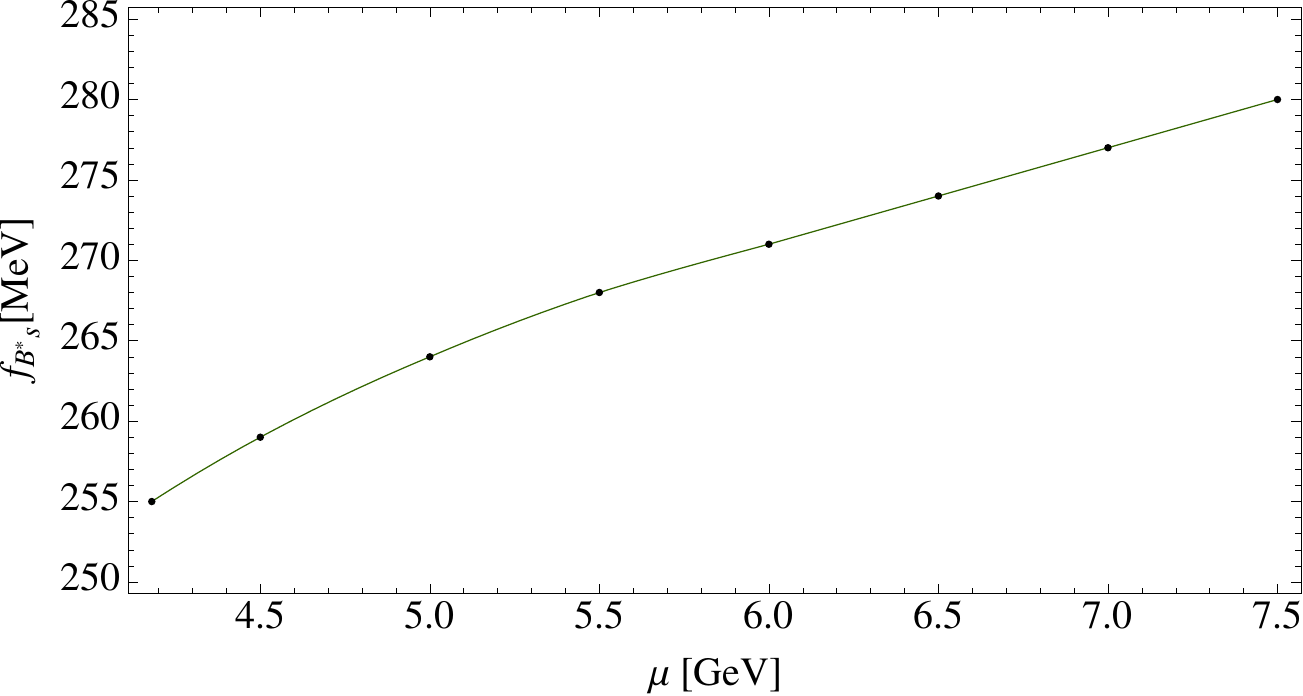}}
\end{center}
\vspace{-0.5cm}
 {\hspace*{0.5cm} a) }\hspace{4.2cm} b)
\caption{
\scriptsize 
 $f_{B^*_s}$: {\bf a)}  $\tau$-behaviour for different $t_c$ at a given 
$\mu=6$ GeV; {\bf b)} $\mu$-behaviour.}
\label{fig:fb*stau} \label{fig:fb*smu}
\end{figure} 
\nin
We show, in Fig. \ref{fig:fb*stau}a, the $\tau$-behaviour of the result for $f_{B^*_s}$  at $\mu$=5.5 GeV and for different values of $t_c$. For $f_{B^*_s}$, $\tau$-stability starts from $t_c\simeq 34$ GeV$^2$ while $t_c$-stability is reached from $t_c\simeq (50-65)$ GeV$^2$.  We show in Fig. \ref{fig:fb*smu}b the $\mu$-behaviour of these optimal results. One can notice a slight inflexion point  at $\mu$=6 GeV which is about the one (5.0-5.5) GeV  obtained previously for the ratio  $f_{B^*_s}/f_{B^*}$. At this value of $\mu$, we obtain:
\bea
f_{B^*_s}&=&271(39)_{t_c}(0)_\tau(3)_{svz}(6)_\mu=271(40)~{\rm MeV}~,\nnb\\
{\rm with}&:&
(3)_{svz}=(1)_{\alpha_s}(1.5)_{\alpha_s^3}(0.5)_{m_b}(2)_{\la\bar dd\ra}(0.5)_{\la \alpha_sG^2\ra}\nnb\\
&&(0.5)_{\la\bar dGd\ra}(0)_{\la g^3G^3\ra}(0.5)_{\la\bar dd\ra^2}(0)_{m_s}(1)\kappa~.
\eea
Combining this result with the one of $f_{B^*}$ in Eq. (\ref{eq:fb*mu}) obtained within the same approach and conditions, we deduce the ratio:
\beq
f_{B^*_s}/f_{B^*}=1.13(25)~,
\lb{eq:rbs1}
\eeq
where the large error is due to the determinations of each absolute values of the decay constants. 
We take as a final value of the ratio $f_{B^*_s}/f_{B^*}$  the most precise determination in Eq. (\ref{eq:rb*stau}).
Combining this result with the final value of $f_{B^*}$  in Eq. (\ref{eq:fb*}) and with the upper bound in Eq. (\ref{eq:fb*_bound}), 
we deduce our final estimate:
\beq
f_{B^*_s}=220(9)~{\rm MeV}~,~~~~f_{B^*_s}\leq 311(19)~{\rm MeV}~.
\lb{eq:fb*s}\lb{eq:fb*s_bound}
\eeq
\section{The scalar meson decay constants $f_{D^*_{0(0s)}}$ and  $f_{B^*_{0(0s)}}$}
\begin{figure}[hbt] 
{\includegraphics[width=4.3cm  ]{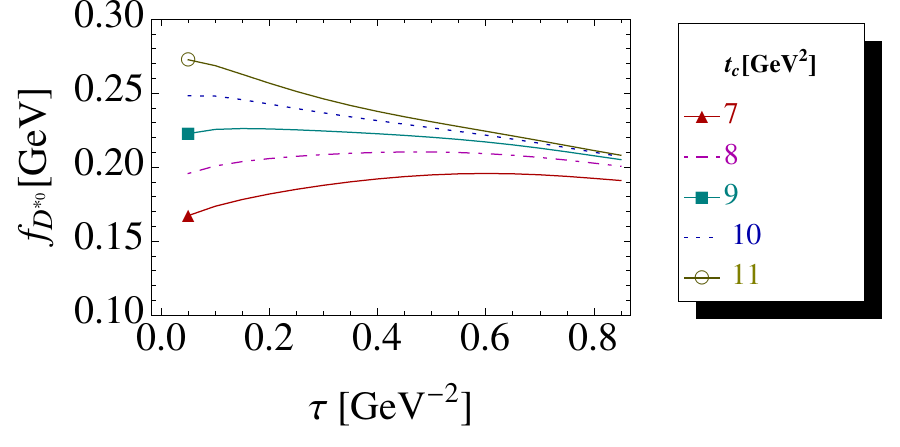}}
{\includegraphics[width=4.3cm  ]{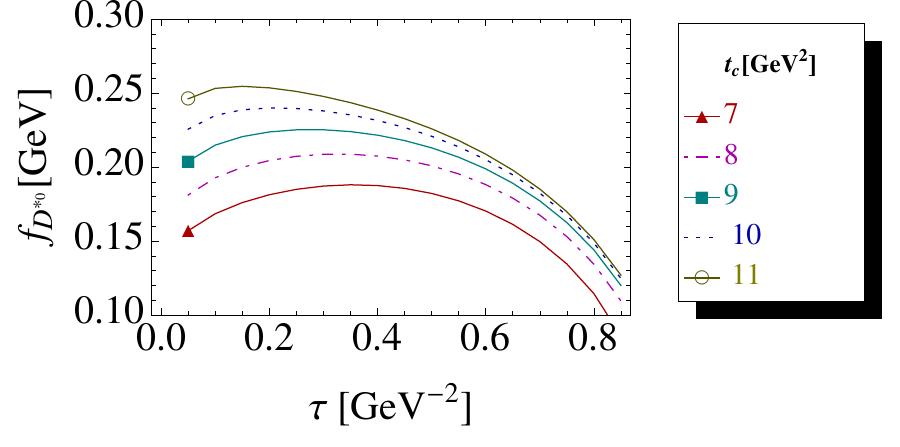}}
\vspace{0.2cm}
 \scriptsize {\hspace*{1cm} a) }\hspace{4.2cm} b)\\
 \vspace{-0.7cm}
 \begin{center}
 {\includegraphics[width=3.5cm  ]{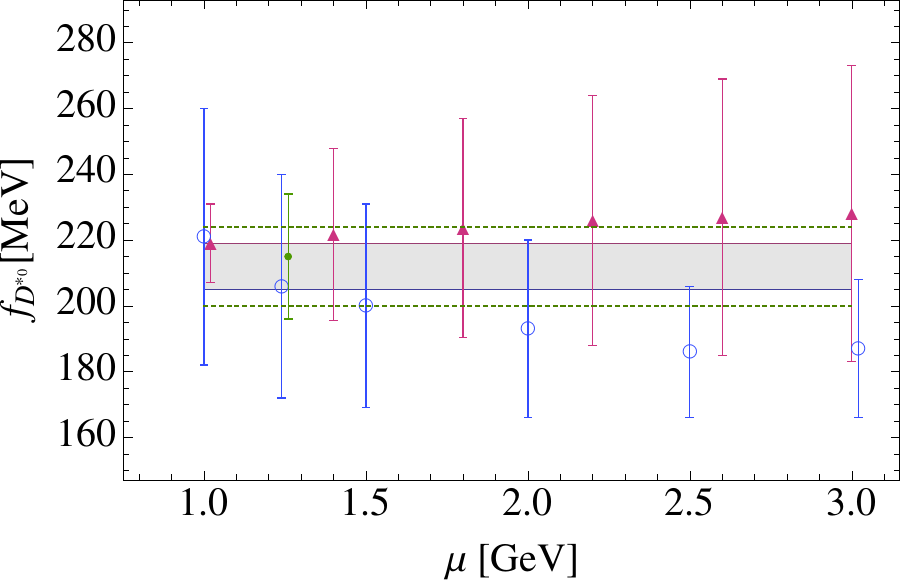}}
{\includegraphics[width=4.6cm  ]{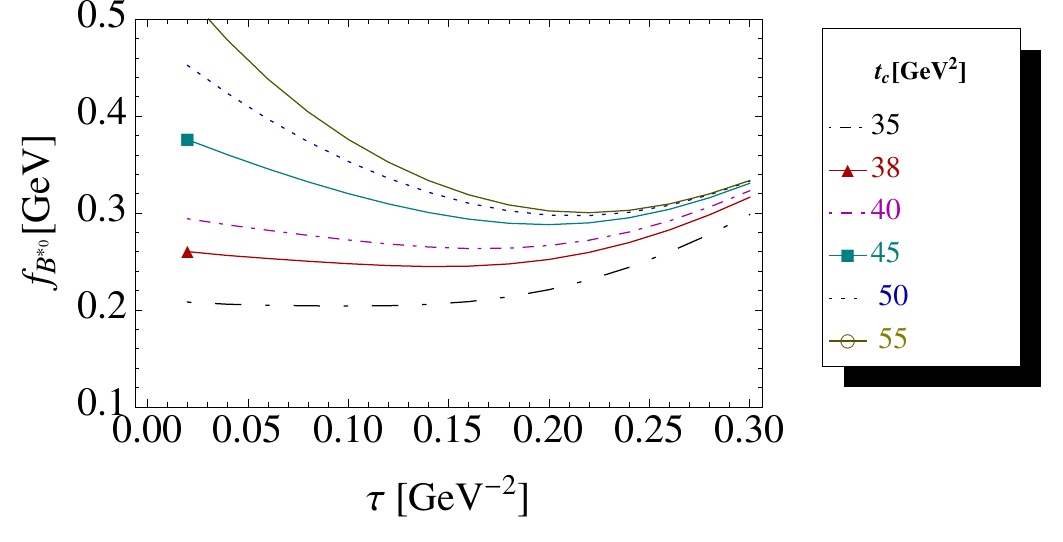}}\\
\end{center}
\vspace{-0.3cm}
 \scriptsize{\hspace*{1cm} c) }\hspace{4.2cm} d)
\caption{
\scriptsize 
{\bf a)} $f_{D^*_0}$  versus $\tau$ and $t_c$ for  $\mu=\tau^{-1/2}$; {\bf b)} The same as a) but for $\mu=$ 2 GeV;  {\bf c)} Optimal value of $f_{D^*_0}$ versus $\mu$: blue open circles are data from SR with running mass. Red triangles are from SR using invariant mass. The green open circle comes from the SR for $\mu=\tau^{-1/2}$. The grey region  is the average with averaged 
errors. The green dotted horizontal lines correspond to the  error coming from the best determination. ;  {\bf d)} $f_{B^*_0}$  versus $\tau$ and $t_c$ for $\mu=1/\sqrt{\tau}$. A similar behaviour is obtained for $f_{B^*_{0s}}$.}
\label{fig:fd*0tau} 
\end{figure}

\begin{figure}[hbt] 
\begin{center}
{\includegraphics[width=4.cm  ]{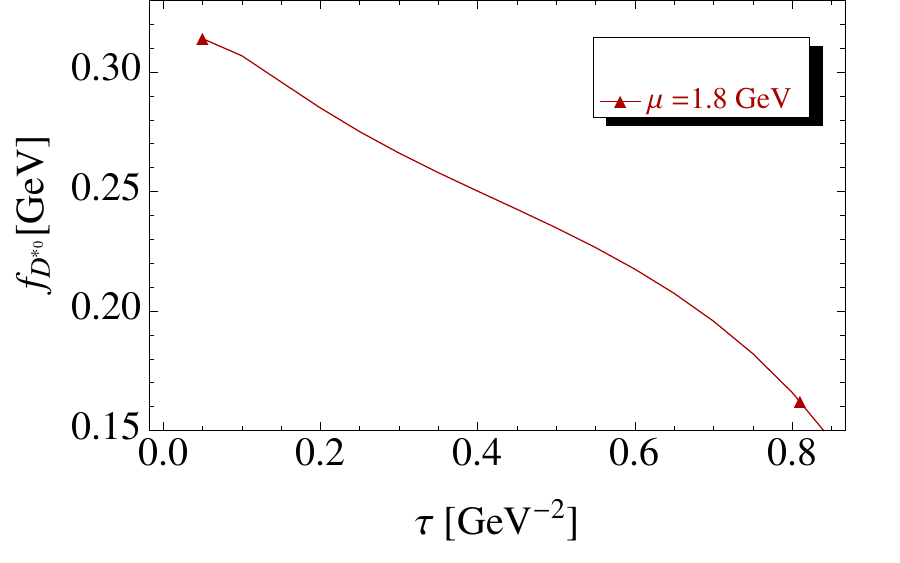}}
{\includegraphics[width=3.6cm  ]{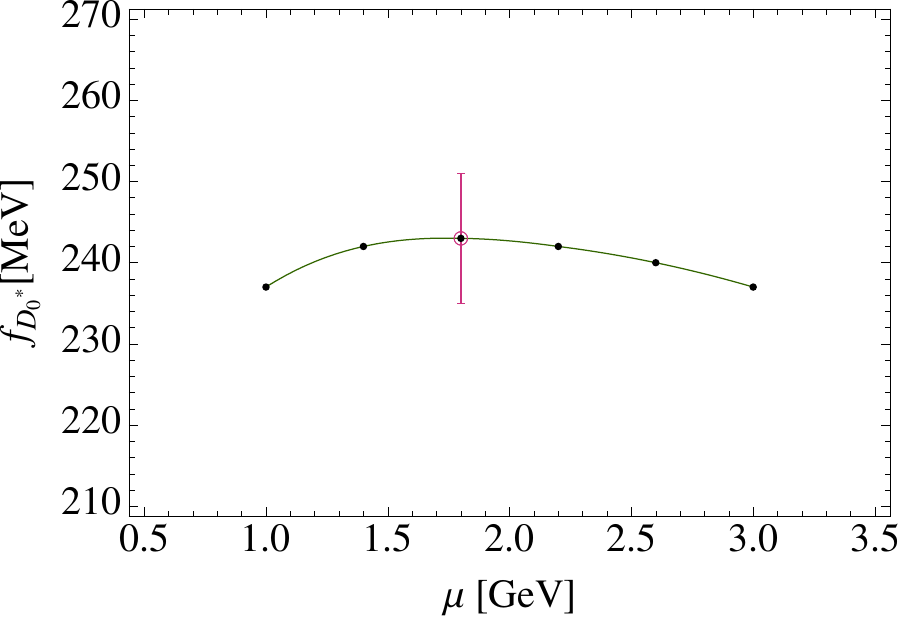}}
\end{center}
\vspace{-0.5cm}
 \scriptsize{\hspace*{1cm} a) }\hspace{4.7cm} b)
\caption{
\scriptsize 
Upper bound of $f_{D^*_0}$: {\bf a)}  versus $\tau$ for  $\mu=1.8$ GeV; {\bf b)} for different $\mu$. The red open cercle is the optimal value.  } 
\label{fig:fd0-bound} 
\end{figure}

\b Due to large discrepancies of the $f_{D^*_{0}}$ values among different SR\,\cite{DAI,COLA2,WANG} and between lattice\,\cite{JUGEAU,HERD} results (see Table \ref{tab:fs}), we update our previous estimates in \cite{SNhl,SNFB88,SNB2}. We use the same QCD inputs (N2LO for PT within the $\overline{MS}$-scheme  $\oplus ~d\leq 6$ condensates) and the same criteria as for $f_{D,B}$\,\cite{SNFB12a} for obtaining the optimal results. The difference beween the scalar and pseudoscalar SR is the overall factor $(M_Q\pm m_s)^2$ and the signs of the $\la\bar qq\ra$ and mixed condensates effects for $m_s=0$.  We show $f_{D^*_0}$ versus $\tau$ in Figs.\,\ref{fig:fd*0tau}a,b where the $\tau$-stability about $(0.4-0.6)$ GeV$^{-2}$ of the SR for $\mu=\tau^{-1/2}$ constrains $t_c$ inside $(7-9$) GeV$^2$ (Fig.\,\ref{fig:fd*0tau}a)\,\footnote{Larger $t_c$ corresponds to an apprent small $\tau$-stability outside the SR window and leads to a larger $f_{D^*_0}$ (Fig. \ref{fig:fd*0tau}b).}. In Fig.\,\ref{fig:fd*0tau}b, we show the SR for a given $\mu=$ 2 GeV, and  $t_c=(7- 11)$ GeV$^2$-stability. The effect of the truncation of the PT series is analyzed by using instead the invariant quark mass in Eq.\,\ref{eq:run} and the QCD scale $\Lambda $. 
We show in Fig.\,\ref{fig:fd*0tau}c the results  versus $\mu$ and using either the running (blue open circle data points)
 or the invariant (red triangle data) mass. Taking their average, we deduce the value in Table\,\ref{tab:fp} where 95\% of the error comes from the above range of $t_c$-values. The error becomes 12 instead of 7 if one considers the most precise determination. Upper bound is derived in Fig. \ref{fig:fd0-bound} using the positivity of the spectral function.  We extract from the SR at $\mu=\sqrt{1/\tau}$ the SU(3) breaking ratio $f_{D^*_{0s}}/f_{D^*_0}$ with the result in Table\,\ref{tab:fs}. The error  of the ratio  comes mainly from the $SU(3)$ breaking terms. These results confirm and improve the previous one in \cite{SNhl}.  
 
\b Doing similar analysis for $f_{B^*_0}$ and $f_{B^*_{0s}}$, we notice that only the SR subtracted at $\mu=\tau^{-1/2}$  (Fig. \ref{fig:fd*0tau}d) exhibits $\tau$-stability for $t_c\simeq (35-38)$ GeV$^2$. We use $M_{B^*_0}\simeq M_{B^*_{0s}}=5733$ MeV assuming that $M_{B^*_0}-M_B\simeq M_{D^*_0}-M_D$ = 448 MeV supported by the SR estimate 413(210) MeV in \cite{SNFB88,SNB2}. We deduce the results in Table\,\ref{tab:fp}, where like for $f_{D^*_0}$, the errors come mainly from $t_c$ and the $SU(3)$ breaking parameters.  These results improve and confirm previous estimates in\,\cite{SNFB88,SNB2}.

\b $f_{D^*_0}$ agrees with the (inaccurate) phenomenological value 206(120)\,MeV from $B\to D^{**}\pi$\,\cite{JUGEAU}. The agreement for $f_{B^*_0}$ but not $f_{D^*_0}$ with the SR one  in \cite{WANG} is puzzling and could be due to a different treatment of the QCD input and evolution parameters which are sensitive at low scale.  A comparison with some other determinations are given in Table\,\ref{tab:fs}. A future precise measurement of $f_{D^*_0}$ will select among the theoretical estimates.
\nin
\vspace*{-0.5cm}
\nin
\section{The axial meson decay constants $f_{D_1}$ and  $f_{B_1}$}
\begin{figure}[hbt] 
{\includegraphics[width=4.3cm  ]{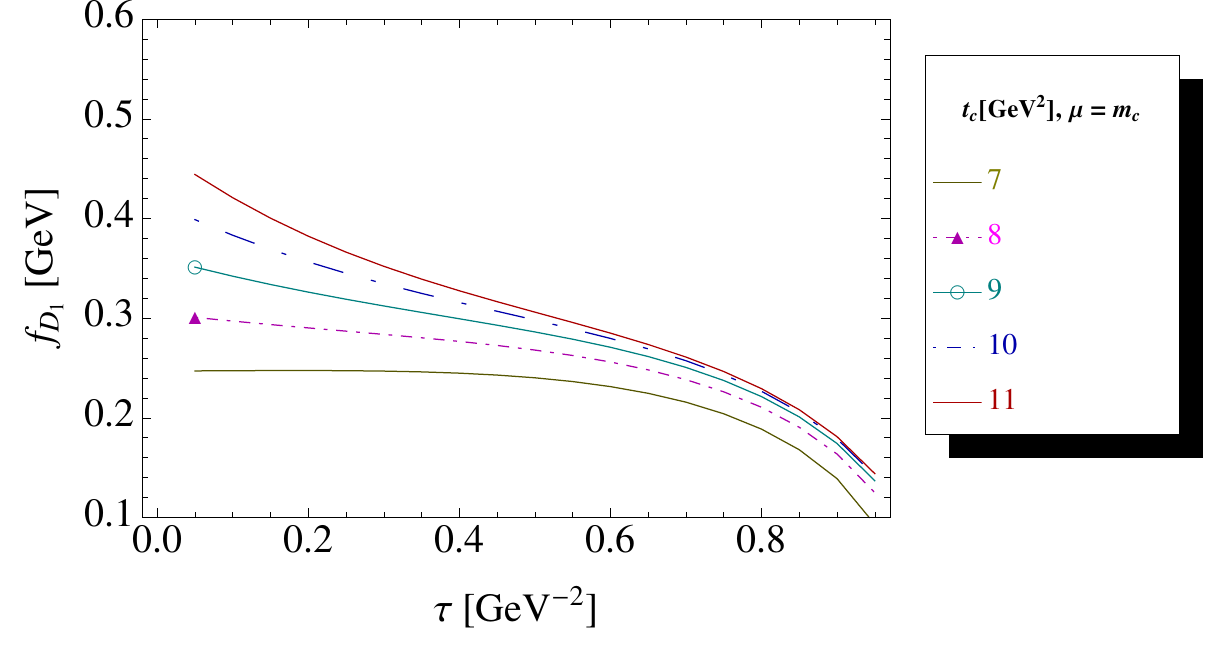}}
{\includegraphics[width=4.3cm  ]{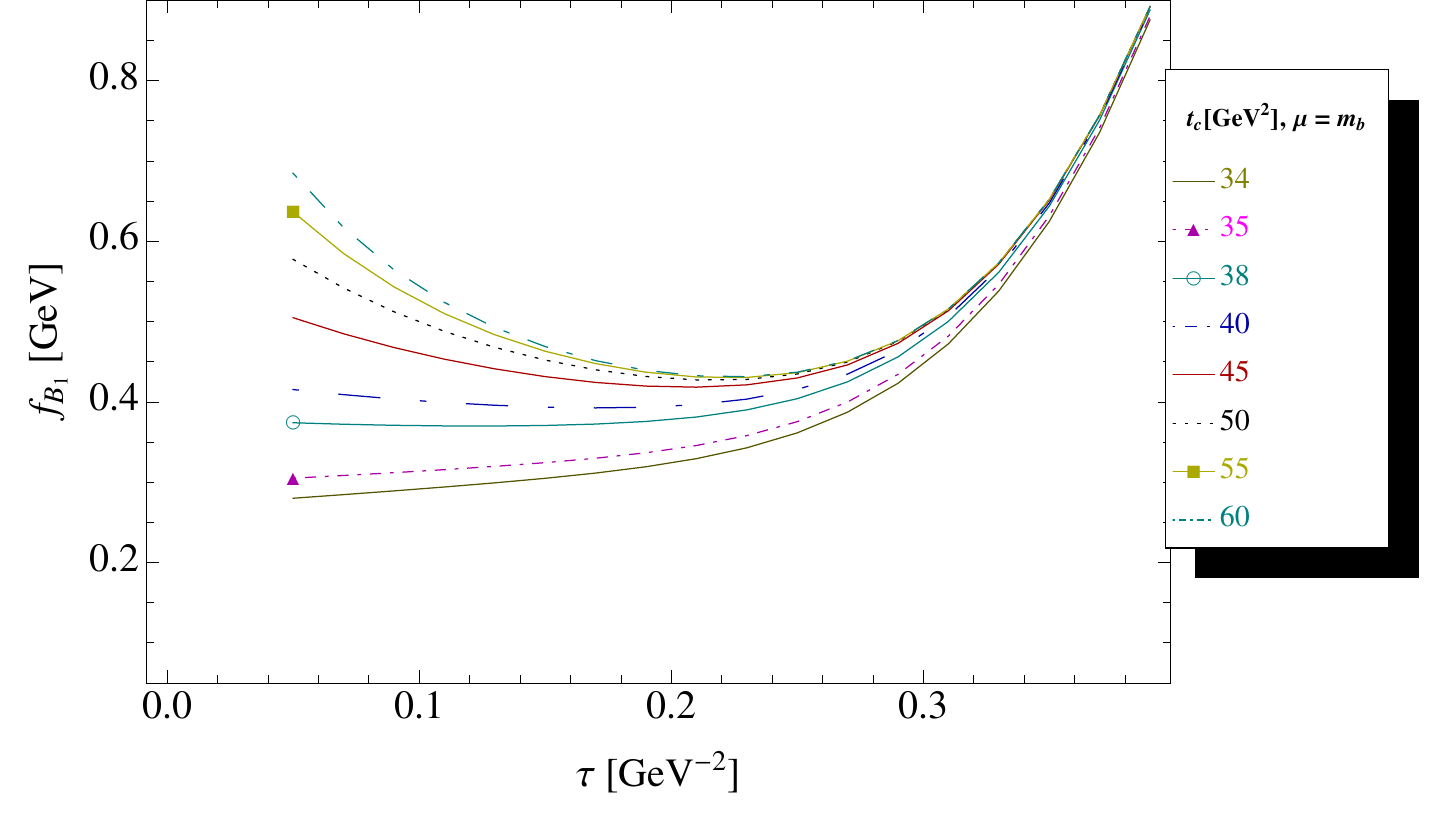}}
\vspace{0.2cm}
 \scriptsize {\hspace*{1cm} a) }\hspace{4.2cm} b)\\
 \vspace{-0.7cm}
 \begin{center}
 {\includegraphics[width=3.6cm  ]{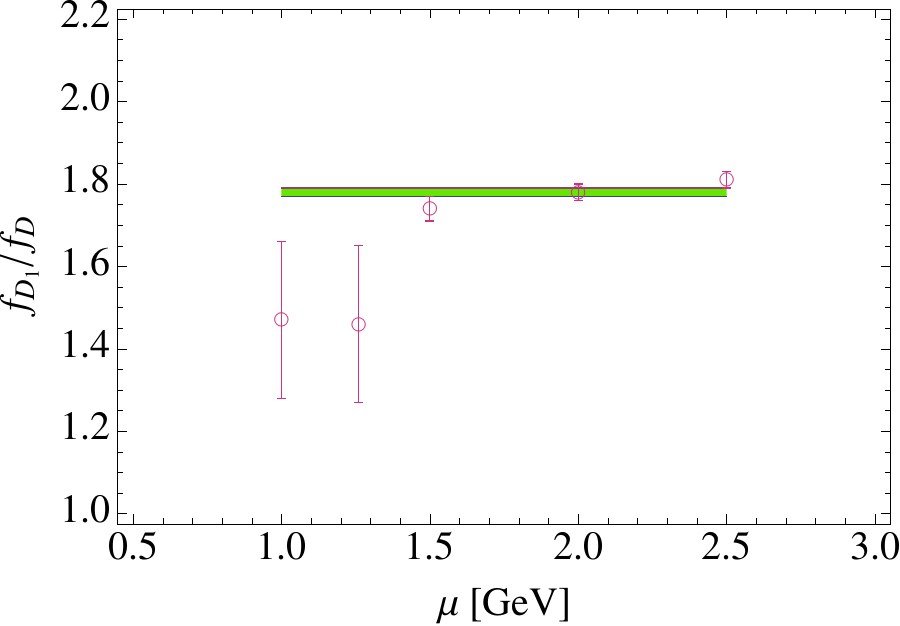}}
{\includegraphics[width=3.6cm  ]{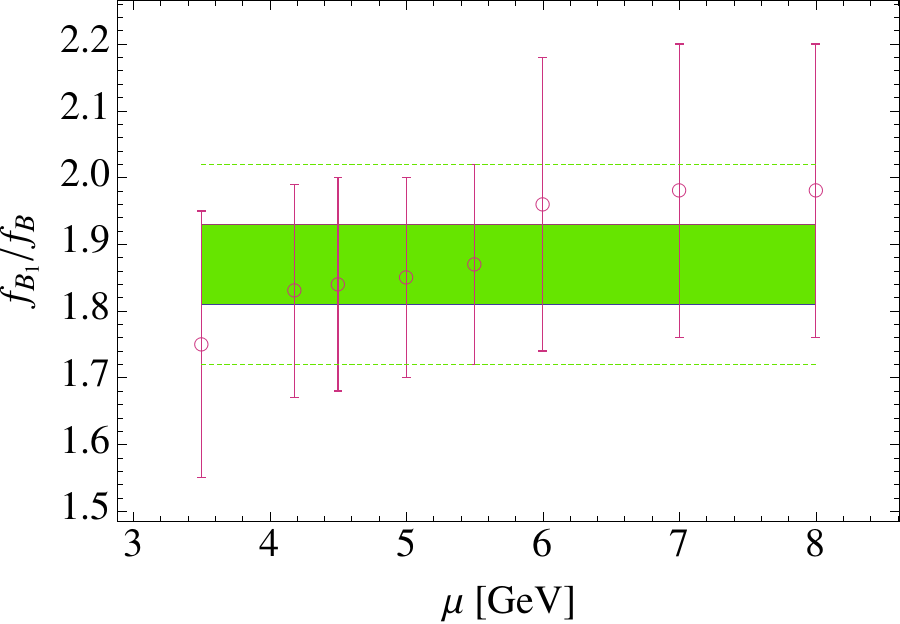}}\\
\end{center}
\vspace{-0.3cm}
 \scriptsize{\hspace*{1cm} c) }\hspace{4.2cm} d)
\caption{
\scriptsize 
{\bf a)} $f_{D_1}$  versus $\tau$ and $t_c$ for  $\mu=m_c$; {\bf b)} The same as a) but for $f_{B_1}$;  {\bf c)} Optimal value of $f_{D_1}$ for different $\mu$: the green dashed line is the average. The green region  is  the value if one 
takes the errors from the weighted average.  ;  {\bf d)} The same as c) but for $f_{B_1}$.  The horizontal green lines correspond to the error from the best determination.}
\label{fig:fd1} 
\end{figure} 
\nin
In the zero  light quark mass limit, the axial-vector correlator can be deduced from the vector one
by changing the sign of the chiral condensate contributions. Using the expression known to N2LO $\oplus$ $d\leq 6$ condensates, we extract directly  the ratio $f_{D_1}/f_D$ and $f_{B_1}/f_B$ by taking each value of $f_{D_1}$ and $f_{B_1}$ at a $\tau$-minimum for a given $(\mu, t_c)$ (Figs.\,\ref{fig:fd1}a,b) and the corresponding value of $f_{D}$ (Fig.\,\ref{fig:fdtau}) and $f_{B}$ (Fig.\,\ref{fig:fbtau}) at the same of set ($\mu, t_c$). We use $M_{B_1}-M_{B^*}\simeq M_{B^*_0}-M_B$=448 MeV as indicated  by the result 417(212) MeV in \cite{SNFB88,SNB2}. The optimal ratios obtained in this way are given for different $\mu$ in Figs.\,\ref{fig:fd1}c,d. The average given in Table\,\ref{tab:res} comes from the dashed coloured regions.
\begin{figure}[hbt] 
\begin{center}
{\includegraphics[width=5cm  ]{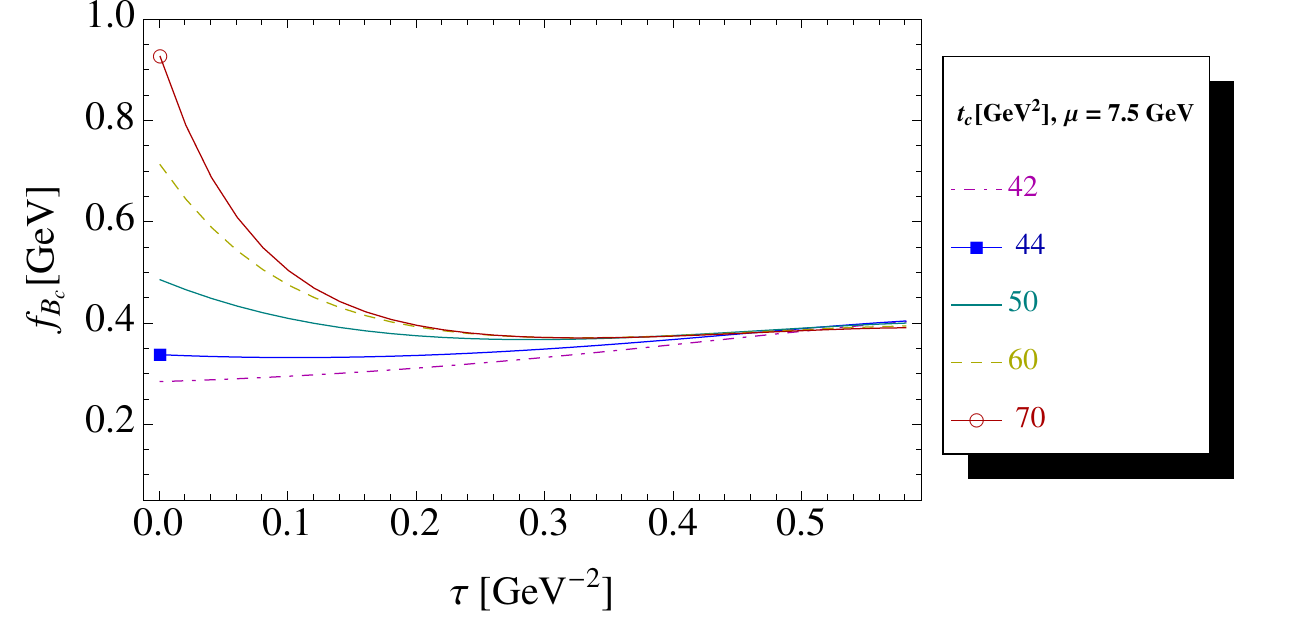}}
\caption{
\scriptsize 
 $\tau$-behaviour of $f_{B_c}$  from  ${\cal L}_{P} $ for different  $t_c$, and for  $\mu=7.5$ GeV. }
\label{fig:fbctau} 
\end{center}
\vspace*{-1cm}
\end{figure} 
\nin
\begin{figure}[hbt] 
\begin{center}
{\includegraphics[width=3.0cm  ]{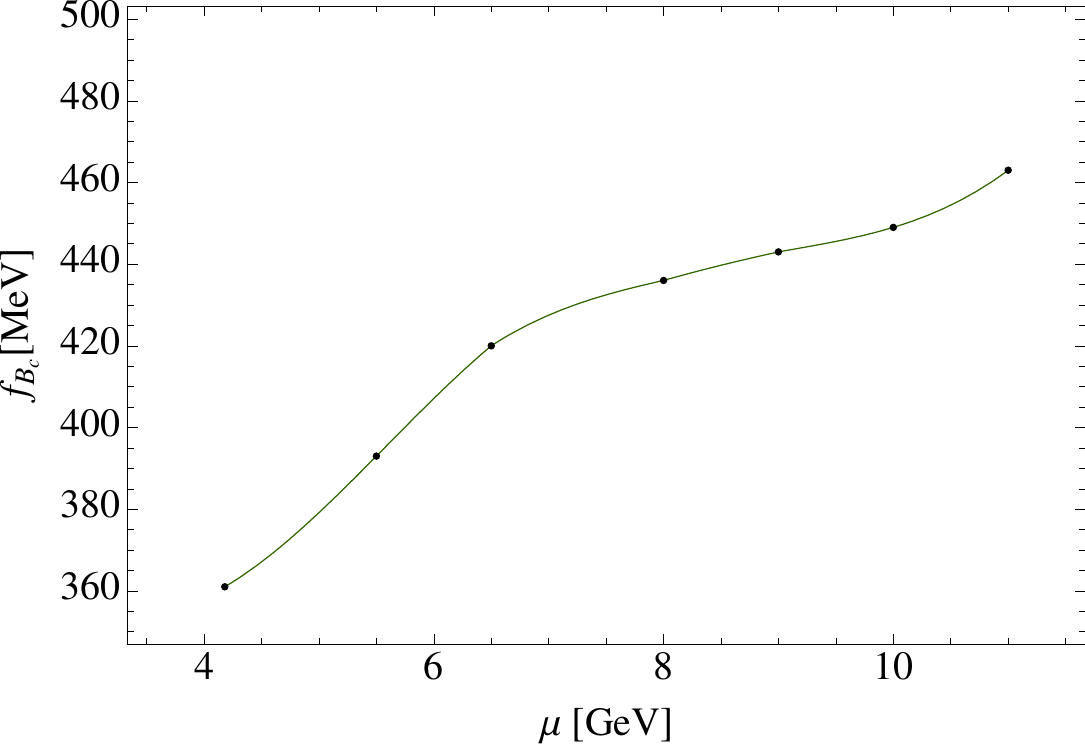}}
{\includegraphics[width=3.8cm  ]{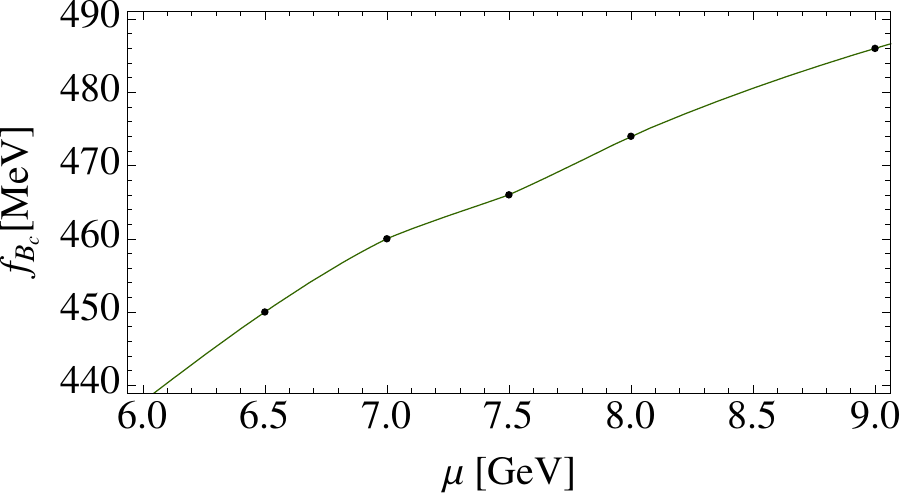}}
\end{center}
\vspace{-0.5cm}
 \scriptsize{\hspace*{1cm} a) }\hspace{3.cm} b)
\caption{
\scriptsize 
$f_{B_c}$  at different values of the subtraction point $\mu$: a) estimate; b) upper bound } 
\label{fig:fbcmu} 
\end{figure} 
\nin
\begin{figure}[hbt] 
\begin{center}
{\includegraphics[width=4.7cm  ]{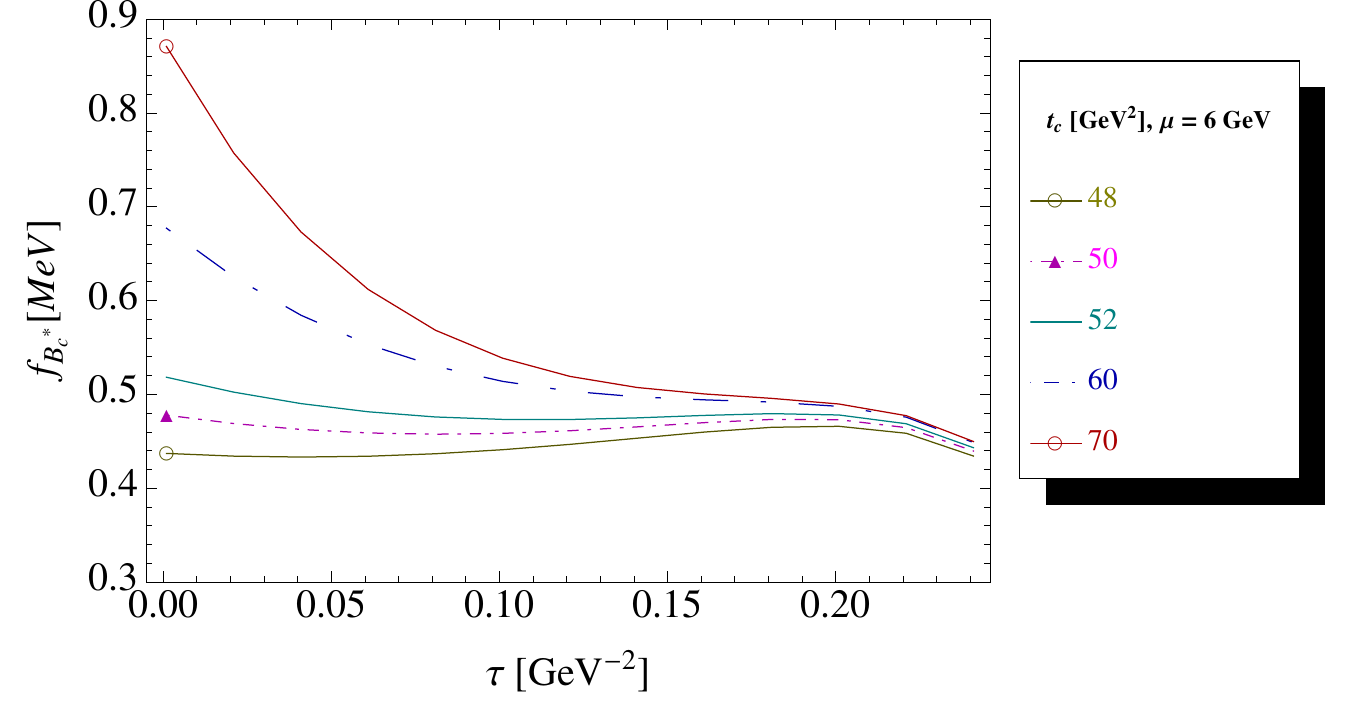}}
{\includegraphics[width=3.5cm  ]{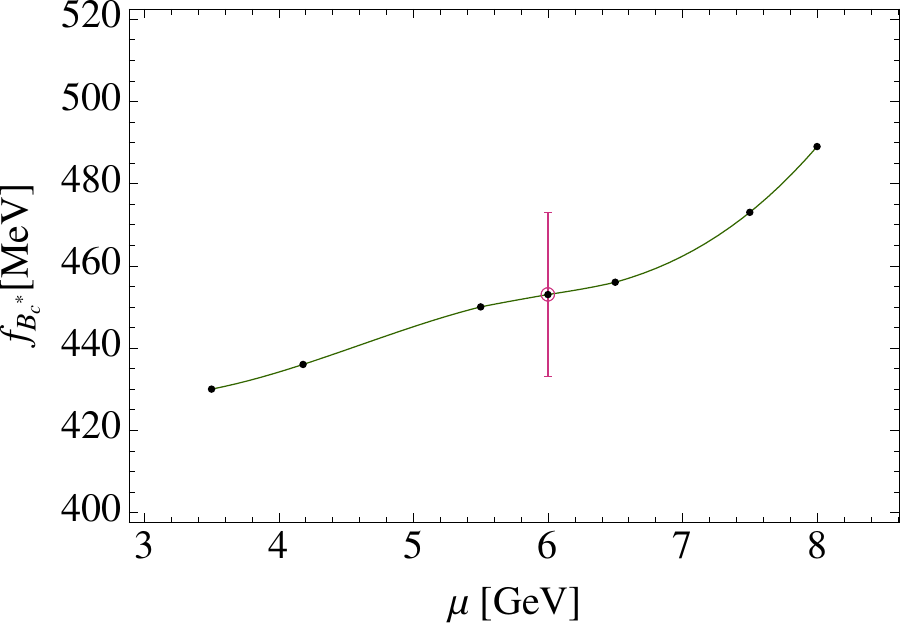}}
\end{center}
\vspace{-0.5cm}
 \scriptsize{\hspace*{1cm} a) }\hspace{4.7cm} b)
\caption{
\scriptsize 
$f_{B^*_c}$ : {\bf a)} versus $\tau$ for different $t_c$ and at a given $\mu$;  {\bf b)} optimal values at $\tau\simeq 0.1$ GeV$^{-2}$  versus $\mu$. he open red circle is the optimal value at the inflexion point. } 
\label{fig:fb*c} 
\end{figure} 
\nin
\begin{figure}[hbt] 
\begin{center}
{\includegraphics[width=4.1cm  ]{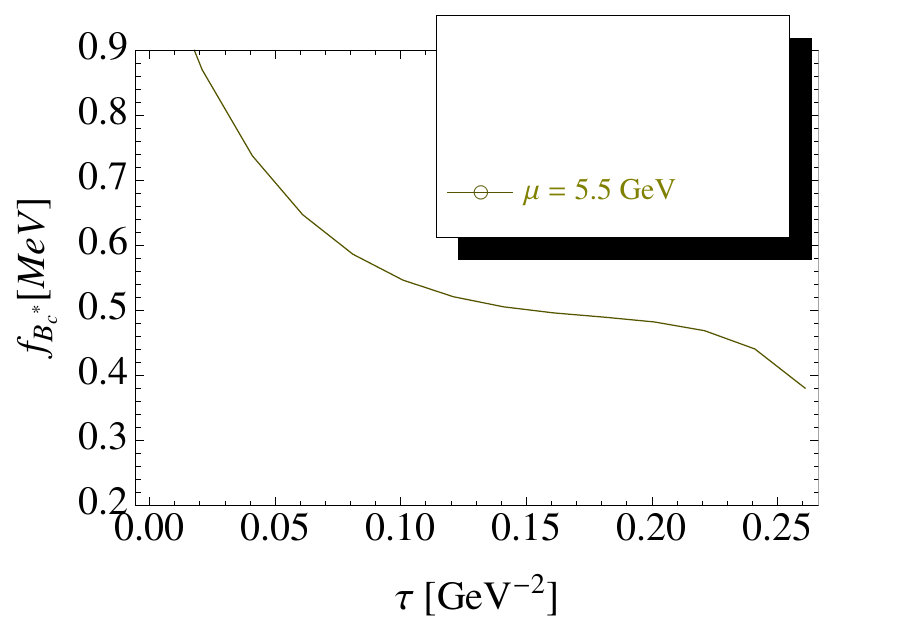}}
{\includegraphics[width=3.9cm  ]{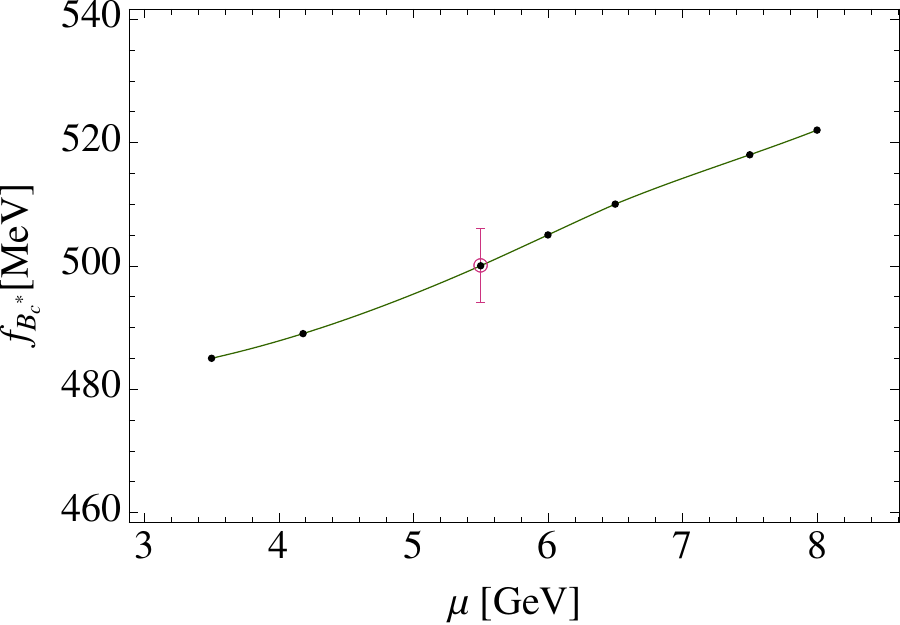}}
\end{center}
\vspace{-0.5cm}
 \scriptsize{\hspace*{1cm} a) }\hspace{4.7cm} b)
\caption{
\scriptsize 
$f_{B^*_c}$ bound : {\bf a)} versus $\tau$ for different $t_c$ and at a given $\mu$;  {\bf b)} optimal values at $\tau\simeq 0.14$ GeV$^{-2}$ versus $\mu$. The open red circle is the optimal bound at the inflexion point. } 
\label{fig:fb*c} 
\end{figure} 
\nin
{\scriptsize
\begin{table}[hbt]
\setlength{\tabcolsep}{0.48pc}
 \tbl{    
  Pseudoscalar $D_{(s)},B_{(s)}$ meson decay constants.  
  NS2R refers to Non-Standard Sum Rules not using either the minimal QCD continuum ansatz nor  the exponential / moments weights. 
  The errors of the Global Average  are rescaled by 1.2 as in \cite{ROSNER} for accounting slight tensions between different determinations. 
  }
    {\scriptsize
\begin{tabular}{lllccc} 
&\\
\hline
\hline
&&&Sources &Refs. \\
\cline{4-5} 
\boldmath$f_{D}$ [MeV] &\boldmath$f_{D_s}$ [MeV]&\boldmath$f_{D_s}/f_{D}$&&\\
204(6)&$246(6)$&1.21(4)&SR & \cite{SNFB12a}\\
203(23)&$235(24)$&1.15(4)&SR & \cite{SNFB3,SNFBREV}\\
$\leq 218(2)$&$\leq 254(2)$&$-$&SR&  \cite{SNFB12a}\\
$\leq 279$&$\leq 321(12)$&$-$&SR& \cite{SNFBREV}\\
$\leq 230$&$\leq 270$&$-$&SR&  \cite{JAMIN3b}\\
201(13)&238(18)&1.15(5)&SR&\cite{PIVOV}\\
208(10)&240(10)&1.15(6)&SR&\cite{WANG}\\
$195(20)$&$-$&$-$&HQET-SR& \cite{PENIN}\\
\bf 204.0(4.6)&\bf 243.2(4.9)&\bf 1.170(23)&\bf SR Average & \\
 209.2(3.3)&249.3(1.8)&1.187(2)& Latt average  $2\oplus 1$& \cite{ROSNER,LATT13}\\
212.1(1.2)&248.8(1.3)&1.172(3)& Latt average  $2\oplus 1\oplus 1$& \cite{ETM14,MILC}\\
\bf 211.8(1.1)&\bf 249.0(1.1)&\bf 1.173(3)& \bf Latt  final average &\\
206(9)&245(16)&1.19(3)&NS2R&\cite{LUCHA1}\\
177(21)&205(22)&1.16(3)&NS2R&\cite{PENA}\\
\bf 211.2(1.1)&\bf 249.1(1.1)&\bf 1.173(3)& \bf Global average &\\
\bf 203.7(4.7)&\bf 257.8(4.1)&$-$&\bf Data & \cite{ROSNER}\\
\cline{1-3}
\\
\boldmath$f_{B}$ [MeV] &\boldmath$f_{B_s}$ [MeV]&\boldmath{$f_{B_s}/f_{B}$} \\
206(7)&$234(5)$&1.14(3)&SR & \cite{SNFB12a}\\
203(23)&$236(30)$&1.16(5)&SR & \cite{SNFB3,SNFBREV}\\
$\leq 235(4)$&$\leq 251(6)$&$-$&SR&  \cite{SNFB12a}\\
197(23)&232(25)&$-$&SR&\cite{JAMIN3a}\\
194(15)&231(16)&1.19(10)&SR&\cite{WANG}\\
207(13)&242(15)&1.17(4)&SR&\cite{PIVOV}\\
$199(29)$&$-$&$-$&HQET-SR&  \cite{SNFB12b}\\
$206(20)$&$-$&$-$&HQET-SR& \cite{PENIN}\\
\bf 204.0(5.1)&\bf 234.5(4.4)&\bf 1.154(21)&\bf SR Average & \\
190(4)&227(2)&1.207(12)&  Latt average  $2\oplus 1$& \cite{ROSNER,LATT13}\\
186(4)&227(4)&1.216(8)& Latt  average $2\oplus 1\oplus 1$&\cite{ETM13,HPQCD13}\\
\bf  188(3)&\bf 227(2)&\bf 1.213(7)&\bf Latt  final average &\\
186(14)&222(12)&1.19(5)&NS2R&\cite{PENA2}\\
192(15)&228(20)&1.18(2)&NS2R&\cite{LUCHA3}\\
\bf  191.9(2.5)&\bf 228.1(1.8)&\bf 1.204(6)&\bf Global average &\\
\bf 196(24)&$-$&$-$&\bf Data & \cite{ROSNER}\\
\\
\hline
\hline
\end{tabular}
}
\label{tab:fp}
\vspace*{-0.25cm}
\end{table}
} 
{\scriptsize
\begin{table}[hbt]
\setlength{\tabcolsep}{0.65pc}
 \tbl{    
 Scalar $D^*_{0(s)},B^*_{0(s)}$ meson decay constants.  
  Values marked with $^\dagger$ are not included in the average. The one with $^*$ has been rescaled by 1.5 for accounting a  slight tension with our estimate.
  }
    {\scriptsize
\begin{tabular}{lllccc} 
&\\
\hline
\hline
&&&Sources &Refs. \\
\cline{4-5} 
\boldmath$f_{D^*_0}$ [MeV] &\boldmath$f_{D^*_{0s}}$ [MeV]&\boldmath{$f_{D^*_{0s}}/f_{D^*_{0}}$}&\\
122(43)$^\dagger$--360(90)$^\dagger$&$-$&$-$&Latt.&\cite{JUGEAU,HERD}\\
128(13)$^\dagger$--373(19)$^\dagger$&$-$&$-$&SR&\cite{DAI,COLA2,WANG}\\
217(24)&$202(23)$&0.93(2)&SR&\cite{SNhl}\\
221(12)&$202(19)$&0.912(23)&SR&New\\
$\leq 243(8)$&$\leq 222(6)$&$-$&SR&New\\
\bf 220(11)&\bf 202(15)&\bf 0.922(15)&\bf SR Average & \\
\cline{1-3}
\\
\boldmath$f_{B^*_0}$ [MeV] &\boldmath$f_{B^*_{0s}}$ [MeV]&\boldmath{$f_{B^*_{0s}}/f_{B^*_{0}}$}& \\
263(51)&$-$&$-$&SR&\cite{SNFB88,SNB2}\\
$281(14)$&274(20)$^*$&$-$&SR&\cite{WANG}\\
271(26)&$233(21)$&0.865(54)&SR&New\\
$\leq 302(8)$&$\leq 261(17)$&$-$&SR&New\\
\bf 278(12)&\bf 255(15)&$-$&\bf SR Average & \\
\\

\hline
\hline
\end{tabular}
}
\label{tab:fs}
\vspace*{-0.25cm}
\end{table}
} 
{\scriptsize
\begin{table}[hbt]
\setlength{\tabcolsep}{0.46pc}
 \tbl{    
Vector $D^*, B^*$ and axial $D_1,B_1$  meson decay constants.  
The  $n_f=2$ Lattice and Global Average errors marked with $^*$ is rescaled by 1.5 for accounting slight tensions between different estimates.
 }
    {\scriptsize
\begin{tabular}{llllccc} 
&\\
\hline
\hline
&&&&Sources &Ref. \\
\cline{5-6}
\boldmath$f_{D^*}$ [MeV]&\boldmath{$f_{D^*}/f_{D}$} &\boldmath$f_{B^*}$ [MeV] &\boldmath{$f_{B^*}/f_{B}$}&  \\
250(11)&$1.218(36)$&209(8)&1.016(16)&SR& \cite{SNFB15}\\
$242(16)$&1.20(8)&210(11)&1.02(5)&SR&\cite{PIVOV}\\
263(21)&$-$&213(18)&$-$&SR&\cite{WANG}\\
$-$&$-$&$-$&1.025(16)&HQET-SR& New \\
$\leq 266(8)$&$-$&$\leq 295(18)$&$-$&SR&\cite{SNFB15}\\
$\leq 297$&$-$&$\leq 261$&$-$&SR&\cite{PIVOV}\\
\bf 250(8)&\bf 1.215(30)&\bf 210(6)&\bf 1.020(11)&\bf SR Average & \\
278(24)$^*$&$1.208(27)$&$-$&$1.051(17)$&Latt. $n_f=2$&\cite{BECIR2}\\
$-$&$-$&$175(6)$&$0.941(26)$&Latt. NR &\cite{DAVIES}\\
$-$&$-$&$-$&\bf 1.018(14)&\bf Latt Average &\\
252(22)&1.22(8)&$182(14)$&$0.944(21)$&NS2R&\cite{LUCHA2}\\
\bf 253(7)&\bf 1.212(20)&\bf 192(6)$^*$&\bf 1.008(12)$^*$&\bf Global Average & \\
\cline{1-4}
\\
\boldmath$f_{D_1}$ [MeV]&\boldmath{$f_{D_1}/f_{D}$} &\boldmath$f_{B_1}$ [MeV] &\boldmath{$f_{B_1}/f_{B}$}&  \\
363(11)&$1.78(1)$&385(18)&1.87(6)&SR&New\\
$\leq 388(5)$&$-$&$\leq 440(16)$&$-$&SR&New\\
333(20)&$-$&335(18)&$-$&SR&\cite{WANG}\\
\bf 356(16)$^*$&$-$&\bf 360(19)$^*$&\bf SR Average & \\


\hline
\hline
\end{tabular}
}
\label{tab:res}
\vspace*{-0.25cm}
\end{table}
} 
{\scriptsize
\begin{table}[hbt]
\setlength{\tabcolsep}{0.33pc}
 \tbl{    
 $SU(3)$ breakings for the vector $D^*_s$ and $B^*_s$ decay constants.
 }
    {\scriptsize
\begin{tabular}{llllcc}
&\\
\hline
\hline
&&&&Sources &Ref. \\
\cline{5-6}
\boldmath$f_{D^*_s}$ [MeV]&\boldmath{$f_{D^*_s}/f_{D^*}$} &\boldmath$f_{B^*_s}$ [MeV] &\boldmath{$f_{B^*_s}/f_{B^*}$}&  \\
270(29)&1.08(6)&220(14)&1.054(11)&SR & \cite{SNFB15}\\
$293(17)$&$1.21(5)$&$210(11)$&$1.20(4)$&SR&\cite{PIVOV}\\
308(21)&$-$&255(19)&$-$&SR&\cite{WANG}\\
$\leq 287(18)$&$-$&$\leq 317(17)$&$-$&SR& \cite{SNFB15}\\
$\leq 347$&$-$&$\leq 296$&$-$&SR&\cite{PIVOV}\\
\bf 290(11)&\bf 1.16(4)&\bf 221(7)&\bf 1.064(10)&\bf SR Average & \\
306(27)&1.21(6)&214(19)&$-$&NS2R&\cite{LUCHA2}\\
311(14)$^*$&1.16(6)&$-$&$-$&Latt. $n_f=2$&\cite{BECIR2}\\
$274(6)$&$-$&$213(7)$&$-$&Latt. $n_f=2\oplus 1$ \& NR &\cite{DAVIES,DAVIES2}\\
\bf 282.5(4.8)&\bf 1.17(3)&\bf 216.9(4.6)&\bf 1.064(11)&\bf Global Average & \\

\hline
\hline
\end{tabular}
}
\label{tab:su3}
\vspace*{-0.5cm}
\end{table}
} 
{\scriptsize
\begin{table}[hbt]
\setlength{\tabcolsep}{0.65pc}
 \tbl{    
Pseudoscalar $B_c$  and vector $B_c^*$ mesons decay constants. The $n_f=2$ Lattice error marked by $^*$ is rescaled by 1.5. The same for NS2R for accounting  the tension with other  estimates. 
 }
{\scriptsize
\begin{tabular}{llllccc} 
&\\
\hline
\hline
&&&&Sources &Ref. \\
\cline{5-6}
\boldmath{$f_{B_c}$} [MeV]&&\boldmath{$f_{B^*_c}$} [MeV]&\boldmath{$f_{B^*_c}/f_{B_c}$}\\
436(40)&$-$&$-$&$-$&SR& \cite{SNFB15}\\
$\leq 466(16)$&$-$&$-$&$-$&SR& \cite{SNFB15}\\
503(171)&$-$&$-$&$-$&Pot. Mod.&\cite{BAGAN}\\
528(29)$^*$&$-$&$-$&$-$&NS2R &\cite{DOM}\\
427(6)&$-$&$-$&$-$&Latt. NR&\cite{LATTFBC}\\
434(23)$^*$&$-$&422(20)$^*$&0.988(27)&Latt. $n_f=2$&\cite{DAVIES}\\
$-$&&453(20)&1.043(45)&SR& New\\
$-$&&$\leq 505(6)$&$-$&SR& New\\
\bf 432(6)&$-$&\bf 438(14)&\bf 1.003(23)&\bf Global Average & \\


\hline
\hline
\end{tabular}
}
\label{tab:fbc}
\vspace*{-0.25cm}
\end{table}
} 
\begin{figure}[hbt] 
\begin{center}
{\includegraphics[width=6.5cm]{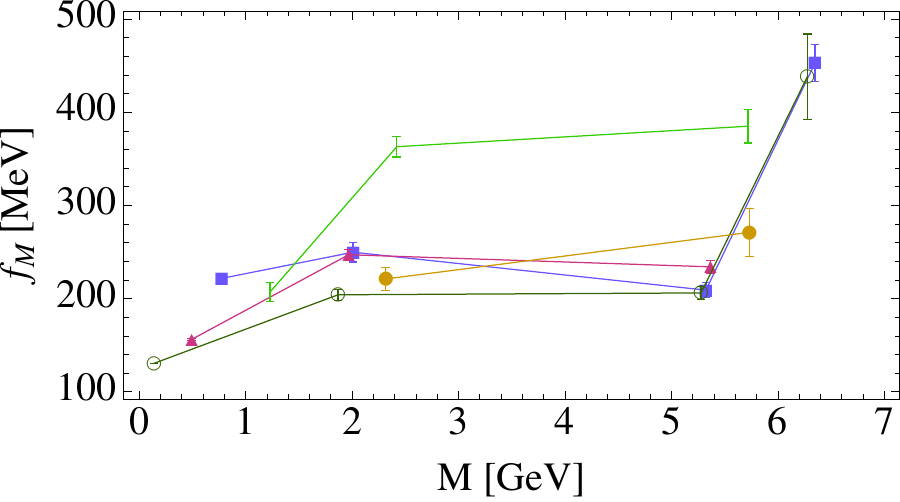}}
\caption{
\scriptsize 
 Behaviour of the meson decay constants versus the meson masses: 
green  open circle:  pseudoscalar $\pi,D,B,{B_c}$ ; purple triangle: $SU(3)$ breaking: $K,{D_{s}},{B_{s}}$; blue box: vector $ \rho,{D^*},{B^*},{B^*_c}$; brown full circle: scalar ${D^*_0}, {B^*_0}$; light green: axial ${A_1},{D_1},{B_1}$. 
 }
\label{fig:fp} 
\end{center}
\vspace*{-0.5cm}
\end{figure} 
\nin
\vspace*{-0.5cm}
\section{The decay constants $f_{B_c}$ and $f_{B^*_c}$}
\nin
We complete the analysis by the estimate and bound of the decay constant $f_{B_c}$ of the $B_c(6277)$ meson $\bar bc$ bound state and of $f_{B^*_c}$ of its vector partner $B^*_c$, where the light quarks $d,s$ are replaced by the heavy quark $c$. Our analysis will be very similar to the one in \cite{SNFB88,BAGAN,CHABAB} but we shall use the running $c$ and $b$ quark masses instead of the pole masses and we shall include N2LO radiative  corrections.

\b The dynamics of the $B_c$ and $B^*_c$ is expected to be different from the $B$ and $B^*$ one because, by using the heavy quark mass expansion, the heavy quark $\la \bar cc\ra$ and quark-gluon mixed $\la \bar cGc\ra$ condensates defined in Section 2  behave as\,\cite{BAGAN}:
\bea
\la \bar cc\ra&=&-{1\over 12\pi m_c}\la\alpha_s G^2\ra-{\la g^3 G^3\ra\over 1440\pi^2 m_c^3}~,\nnb\\
\la \bar cGc\ra&=&{m_c\over \pi}\ga\log{m_c\over \mu}\dr\la\alpha_s G^2\ra-{\la g^3 G^3\ra\over 48\pi^2 m_c}~.
\eea
These behaviours are in contrast with the ones of the light quark  $\la\bar qq\ra$ and mixed quark-gluon $\la \bar qGq\ra$ condensates\,\cite{SNB1,SNB2}.

\b The complete expression of the perturbative NLO pseudoscalar spectral function has been obtained in \cite{GENERALIS} and explicitly written in \cite{BAGAN}, where we transform the pole to the running masses. We add to this expression the N2LO result in \cite{CHET} for $m_c=0$. We consider as a source of errors the N3LO contribution from an assumed geometric growth of the PT series \cite{NZ} which mimics the $1/q^2$-term parametrizing the large order terms of PT series\,\cite{CNZ,ZAK}. 

\b The Wilson coefficients of the non-perturbative $\la \alpha_s G^2\ra$ and $\la g^3 G^3\ra$ contributions are also given in \cite{BAGAN}. 

\b Like in the previous cases, we study the SR versus $\tau$ and $t_c$ as  shown in Fig. \ref{fig:fbctau}. One can note that the non-perturbative contributions are small (about $1-2$ MeV) indicating that the dynamics of the $B_c$ meson is dominated by the perturbative contributions. This feature might explain the success of the non-relativistic potential models for describing the $B_c$-like hadrons \cite{BAGAN}. The optimal result is obtained from $t_c=44$ GeV$^2$ (beginning of $\tau$-stability) until $t_c=(50-60)$ GeV$^2$ (beginning of $t_c$-stability). 
We show in Fig. \ref{fig:fbcmu} the $\mu$-behaviour of the results, where there is an inflexion point for $\mu=(7.5\pm 0.5)$ GeV for the estimate (Fig. \ref{fig:fbcmu}a) and the upper bound (Fig. \ref{fig:fbcmu}b).  At these optimal points, we deduce:
\bea
f_{B_c}&=&436(38)_{t_c}(2)_{\alpha_s}(2)_{\alpha_s^3}(7)_{m_c}(6)_\mu
=436(40)~{\rm MeV}~,\nnb\\
f_{B_c}&\leq& 466(9)_{\alpha_s}(2)_{\alpha_s^3}(12)_{m_c}(8)_\mu=466(16)~{\rm MeV}~.
\label{eq:fbc_bound}\label{eq:fbc}
\eea
We may consider these results as  a confirmation and improvement of the earlier ones  in \cite{SNB2,SNFB88,BAGAN,CHABAB}. 
The results in Eq. (\ref{eq:fbc}) will restrict the wide range of $f_{B_c}$ 
given in the current literature 
and may be used for extracting the CKM angle $V_{cb}$ from the $B_c \to \tau \nu_\tau$ leptonic width.

\b We do a similar analysis for the $B^*_c$ vector meson where we replace the strange by the charm quark mass in the $B^*_s$ SR expression\,\footnote{A complete expression of the spectral function to NLO is given in \cite{RRY} but it has singularity for $m_c\to 0$ and needs to be checked.}. We use the rescaled $M_{B^*_c}=6350(20) $ MeV from potential model \cite{BAGAN}  using the experimental value of $M_{B_c}$. The different steps are shown in Fig.\,\ref{fig:fb*c}. We obtain the results in Table\,\ref{tab:fbc} for $\tau\simeq 0.15$ GeV$^{-2}$. 
\section*{Summary and Conclusions}

\b We have reviewed our recent determinations of the heavy-light pseudoscalar \cite{SNFB12a,SNFB12b}, vector \cite{SNFB15} mesons decay constants from the standard Laplace and Moment sum rules (SR) introduced by SVZ \cite{SVZ}. Motivated by different tensions in the literature, we have completed the review by presenting new determinations of the vector  $f_{B^*}/f_B$,
scalar  $f_{D^*_0, B^*_0}$, axial $f_{D_1,B_1}$ and  vector $f_{B^*_c}$ meson decay constants. 
Our results based on stability criteria are summarized in Tables \ref{tab:fp}--\ref{tab:fbc} and compared with some other recent sum rules and lattice results. Our quoted errors are mainly due to the choice of the QCD continuum threshold and to the estimate of the $\alpha_s^3$ PT series. This latter is not considered in some other SR results. 

\b Then, we have attempted to
give a Global Average of sum rules and lattice results (results from some other approaches can e.g be found in\,\cite{HWANG}), which can be used for further phenomenological applications.
To take into account 
the slight tensions among different determinations, we have rescaled some errors.

\b One can see in these Tables that there are fair agreements among estimates from different standard sum rule (SR) approaches. The slight difference is mainly due to a different appreciation of the optimal results from the choice of the QCD continuum threshold, subtraction constant $\mu$ and of the set of input QCD parameters. The quoted upper bounds come from the positivity of the spectral functions.

\b A slight tension exists between SR, NS2R and lattice results for  $f_{B^*}/f_B$ \cite{LUCHA2}.  To give insight into this problem, we have rederived this ratio in Section 8 from HQET sum rules.  There is also a tension for $f_{B_c}$ from NS2R\,\cite{DOM} (see comments in \cite{SNFB15}).  The N2SR results do not affect significantly the SR average and they are considered in the Global Average.

\b Large tensions exist in the scalar sector among various SR and between lattice results for $f_{D^*_0}$. Our updated  results given in Table\,\ref{tab:fs} using the running $\overline{MS}$ charm mass and SR stability criteria confirm our previous findings\,\cite{SNhl}. 


\b The results indicate a good realization of heavy quark symmetry for the $B$ and $B^*$ mesons ($f_B\approx f_{B^*}$) as expected from HQET \cite{NEUBERT2} but signal  large charm quark mass and radiative QCD corrections for the $D$ and $D^*$ mesons ($f_D\approx f_B,..$) which are known since a long time (see e.g among others \cite{SNFB,SNIMET}). 

\b The $SU(3)$ breaking is typically 20\% for $f_{D_s}/f_D\approx f_{B_s}/f_B$ and 10\% for  $f_{D^*_s}/f_{D^*}\approx f_{B^*_s}/f_{B^*}$ which are almost constant  as they behave like $m_s/\omega_c$\,\cite{SNFBSU3} where $\omega_c\equiv \sqrt{t_c}-M_b\approx M_B'-M_B$  is independent of $M_b$. The reverse effect for $f_{D^*_s}/f_{D^*}\approx f_{B^*_s}/f_{B^*}< 1$ comes mainly from the overall $(1-m_s/m_Q)$ factor not compensated by $M_{D^*_{0s}/B^*_{0s}}\simeq M_{D^*_0/B^*_0}$\,\cite{SNFBSU3}. 

\b It is informative to show the behaviour of our predictions of the pseudoscalar and vector meson decay constants versus the corresponding meson masses in Fig. \ref{fig:fp}. 
We use $f_\rho=(221.6\pm1.0)$ MeV from its electronic width\,\cite{PDG} and $f_{A_1}=207$ MeV from SR analysis\,\cite{SNB2}. One can notice similar $M_Q$ behaviours of these couplings where the 
the $1/\sqrt{M_Q}$ HQET relations\,\cite{HQET,NEUBERT2} are not satisfied due to large $\alpha_s$ and $1/M_Q$ corrections as noted  earlier \,\cite{SNFB,SNIMET}. One can notice small $SU(3)$ but large $SU(4)$ breaking for $B_c,B^*_c$. Chiral symmetries are badly broken between the vector $D^*,B^*$ and axial $D_1,B_1$ meson couplings. 

\b We do not found any deviation of these Standard Model results from the present data. 

\vspace*{-0.25cm}


\begin{thebibliography}{999}
\bibitem{SNFB12a}S. Narison, {\it Phys.Lett.} {\bf B718} (2013) 132;
S. Narison, {\it Nucl. Phys. Proc. Suppl.} {\bf  234} (2013) 187.
\bibitem{SNFB12b}S. Narison, {\it Phys.Lett.} {\bf B721} (2013) 269.
\bibitem{SNFB15}S. Narison, {\it  Int. J. Mod. Phys. } {\bf A30} (2015) 20, 1550116.

  
\bibitem{SNB2}S. Narison, {\it 
World Sci. Lect. Notes Phys.} {\bf 26}
(1989) 1-526.

\bibitem{SNFDTEST} S. Narison,  {\it Phys. Lett.}  {\bf B668} (2008) 308.

\bibitem{MAHMOUDI}A. G. Akeroyd and F. Mahmoudi, {\it JHEP} {\bf 0904} (2009) 121; A. G. Akeroyd and C. H. Chen, {\it Phys. Rev.} {\bf D75} (2007) 075004.

\bibitem{GRINSTEIN} B. Grinstein and J.M Camalich, arXiv: 1509.05049 (2015).

\bibitem{BECIR2}D. Becirevic et al., arXiv: 1407.1019 [hep-ph] (2014);    
D. Becirevic et al., {\it JHEP} {\bf 1202} (2012) 042.

\bibitem{ROSNER}J. Rosner, S. Stone and R.S  Van de Water,  arXiv:1509.02220 (2015).




\bibitem{SVZ} M.A. Shifman, A.I. and Vainshtein and V.I. Zakharov,
{\it Nucl. Phys.} {\bf B147} (1979) 385; ibid, {\it Nucl. Phys.} {\bf B147} (1979) 448.
\bibitem{SNB1} S. 
Narison, {\it 
Cambridge Monogr. Part. Phys. Nucl. Phys. Cosmol.} {\bf 17} (2002) 1-779.
[hep-ph/0205006].

\bibitem{SNB3}S. Narison, {\it Phys. Rept.} {\bf 84} (1982) 263; S. Narison, {\it Acta Phys. Pol.} {\bf B26} (1995) 687; S. Narison, hep-ph/9510270 (1995).

\bibitem{RRY} L.J. Reinders, H. Rubinstein and S. Yazaki, {\it Phys. Rept. }
{\bf 127}  (1985) 1. 

\bibitem{CK}E. de Rafael, hep-ph/9802448.


\bibitem{SNZ}S. Narison, {\it Z. Phys.} {\bf C14} (1982) 263. 

\bibitem{GENERALIS}S.C. Generalis, Ph.D. thesis, Open Univ. report, OUT-4102-13 (1982).

\bibitem{SNFB}S. Narison, {\it Phys. Lett.}  {\bf B198} (1987)  104; S. Narison, {\bf B285} (1992) 141.

\bibitem{SNFB88}S. Narison, {\it Phys. Lett.}  {\bf B210} (1988)  238. 

\bibitem{SNFB3}S. Narison, {\it Phys. Lett.}  {\bf B520} (2001)  115. 

 \bibitem{SNFBREV} S. Narison,  arXiv: hep-ph/0202200(2002)  and references therein.

\bibitem{JAMIN3a}M. Jamin and B. O. Lange, {\it Phys. Rev.} {\bf D65} (2002) 056005.

\bibitem{SNhl}S. Narison, {\it Phys. Lett.} {\bf B605} (2005) 319.

\bibitem{JAMIN3b}A. Khodjamirian, {\it Phys. Rev.} {\bf D79} (2009) 031503; 

\bibitem{JAMIN3c}D.J. Broadhurst and M. Grozin,  {\it Phys. Lett.}  {\bf B274} (1992)  421; 
E. Bagan, P. Ball, V. Braun and H.G. Dosch,  {\it Phys. Lett.}  {\bf B278} (1992)  457; 
 V. Eletsky and A.V. Shuryak,  {\it Phys. Lett.}  {\bf B276} (1993)  365; 
 
 
 \bibitem{NEUBERT2}M. Neubert, {\it Phys. Rept.} {\bf 245} (1994) 259; M. Neubert, {\it Phys. Rev.} {\bf D45} (1992) 2451; ; private communication from Matthias Neubert.

  
  \bibitem{BALL} P. Ball,  {\it Nucl. Phys. }  {\bf B421} (1994), 593.
 
  \bibitem{PENIN}A. Penin and and M. Steinhauser, {\it Phys. Rev.} {\bf D65} (2002) 054006.

\bibitem{NSV2Z}V.A. Novikov et al., 8th conf. physics and neutrino astrophysics (Neutrinos 78),
Purdue Univ. 28th April-2nd May 1978 [see also :  S.S Gershtein and M. Yu Khlopov, {\it JETP Lett.} {\bf 23} (1976) 338].




 \bibitem{SNRAF} S. Narison, E. de Rafael, {\it Phys. Lett.} {\bf B103} (1981) 57.

  \bibitem{SNmass} S. Narison, {\it  Phys.Rev.} {\bf D74} (2006) 034013; 
 S. Narison, {\it Phys.Lett.} {\bf B466} (1999) 345.
 
\bibitem{BNP}E. Braaten, S. Narison and A. Pich, {\it Nucl. Phys.} {\bf B373} (1992) 581;
S. Narison and A. Pich, {\it Phys. Lett.} {\bf  B211} (1988) 183.

\bibitem{SNFBSU3}S. Narison, {\it Phys. Lett.} {\bf B322} (1994) 247.

\bibitem{BROAD} D.J. Broadhurst, S.C. Generalis, Open Univ. rep., OUT-4102-8/R (1982).

\bibitem{BROAD1}D.J. Broadhurst, {\it Phys. Lett.}  {\bf B101} (1981)  423 and private communication.

\bibitem{CHET}K.G. Chetyrkin, M. Steinhauser,  {\it Phys. Lett.}  {\bf B502} (2001)  104.

\bibitem{JAMIN}M. Jamin,  M. M\"unz, {\it Z. Phys.} {\bf C60} (1993) 569.

\bibitem{NZ}S. Narison, V.I. Zakharov, {\it Phys. Lett.} {\bf B679} (2009) 355.

\bibitem{CNZ} K.G. Chetyrkin, S. Narison, V.I. Zakharov, {\it Nucl. Phys.} 
{\bf B550} (1999)  353;
S. Narison, V.I. Zakharov, {\it  Phys. Lett.} {\bf B522} (2001) 266.

\bibitem{ZAK} For reviews, see e.g.: V.I. Zakharov, {\it Nucl. Phys. Proc. Suppl.} 
{\bf 164} (2007) 240; S. Narison,  {\it Nucl. Phys. Proc. Suppl.} {\bf 164} 
 (2007) 225.


\bibitem{TAR}R. Tarrach, {\it Nucl. Phys.} {\bf B183} (1981) 384.

\bibitem{COQUE} R. Coquereaux, {\it Annals of Physics} {\bf 125} (1980) 401;
P. Binetruy and T. S\"ucker, {\it Nucl. Phys.} {\bf B178} (1981) 293.

\bibitem{SNPOLE} S. Narison, {\it  Phys. Lett.} {\bf B197} (1987) 405; S. Narison, 
{\it  Phys. Lett.} {\bf B216} (1989) 191.

\bibitem{BROAD2} N. Gray, D.J. Broadhurst, W. Grafe, and K. Schilcher, {\it Z. Phys.} {\bf  C48} (1990) 673;
J. Fleischer et al., {\it Nucl. Phys.} {\bf B539}
(1999) 671.

\bibitem{CHET2}K.G. Chetyrkin and M. Steinhauser, {\it Nucl. Phys.} {\bf B573}
(2000) 617; K. Melnikov and T. van Ritbergen, hep-ph/9912391; K.G. Chetyrkin,  J.H. K\"uhn and M. Steinhauser, hep-ph/0004189.


\bibitem{SNTAU}S. Narison, {\it Phys. Lett.} {\bf B673} (2009) 30.

\bibitem{JAMI2}Y. Chung et al.{\it Z. Phys.} {\bf C25} (1984)  151;  
H.G. Dosch, M. Jamin and S. Narison, {\it Phys. Lett.} {\bf B220} (1989)  251.

\bibitem{LNT}G. Launer, S. Narison and R. Tarrach, {\it  Z. Phys.} {\bf C26}
(1984) 433.







\bibitem{BETHKE} See e.g:  S. Bethke, {\it Chin. Phys. } {\bf C38} (2014) 090001.


\bibitem{PDG} K.A Olive et al. (PDG), {\it Phys. Rev.} {\bf D86} (2012) 010001.

\bibitem{SNH10}S. Narison,  {\it Phys. Lett.} {\bf B693} (2010)  559; Erratum ibid 705 (2011) 544;
ibid, {\it Phys. Lett.} {\bf B706} (2011)  412; ibid, {\it Phys. Lett.} {\bf B707} (2012)  259. 

\bibitem{IOFFE} B.L. Ioffe and K.N. Zyablyuk, {\it Eur. Phys. J.} {\bf  C27}
(2003)  229 ; B.L. Ioffe, {\it Prog. Part. Nucl. Phys.} {\bf 56} (2006) 232.

 \bibitem{SNmass98}H.G. Dosch and S. Narison,  {\it Phys. Lett.}  {\bf B417} (1998) 173; S. Narison,  {\it Phys. Lett.} {\bf B216} (1989) 191.
 
 \bibitem{SNLIGHT}S. Narison,  {\it Phys. Lett.}  {\bf B738} (2014) 346.
  
\bibitem{HEID}B.L. Ioffe, {\it Nucl. Phys.} {\bf B188} (1981)  317; B.L. Ioffe, {\it Nucl. Phys.} {\bf
B191} (1981) 591; A.A.Ovchinnikov and A.A.Pivovarov,
{\it Yad.\ Fiz.}  {\bf 48} (1988) 1135.


\bibitem{SNI}S. Narison, {\it Phys. Lett.} {\bf B300} (1993) 293; {\it Phys. Lett.} {\bf B361} (1995) 121.

\bibitem{SNGH}S. Narison, {\it Phys. Lett.} {\bf B387} (1996) 162.
\bibitem{YNDU}F.J. Yndurain, hep-ph/9903457.
\bibitem{SNHeavy}S. Narison, {\it Phys. Lett.} {\bf B387} (1996) 162.

\bibitem{SNG1} S. Narison, {\it Phys. Lett.} {\bf B361} (1995) 121; S. Narison,  {\it Phys. Lett.} {\bf B624} (2005) 223.

\bibitem{BELL}J.S. Bell and R.A. Bertlmann, {\it Nucl. Phys.} {\bf B227} (1983) 435;
R.A. Bertlmann, {\it Acta Phys. Austriaca} {\bf 53} (1981) 305 and references therein.
 
 \bibitem{SNTAU9} S. Narison, {\it Phys. Lett.} {\bf B673} (2009) 30 and references therein.
 
  \bibitem{HBARYON} R.M. Albuquerque, S. Narison, {\it Phys. Lett.} {\bf B694} (2010) 217; R.M. Albuquerque, S. Narison, M. Nielsen, {\it Phys. Lett.} {\bf B684} (2010) 236.
  
  \bibitem{MCNEILE}  C. McNeile et al., {\it Phys. Rev.} {\bf D87} (2013) 034503.
  
   \bibitem{FNR}E.G. Floratos, S. Narison and E. de Rafael, {\it Nucl. Phys.} 
{\bf B155} (1979) 155.

 \bibitem{LATT13}S. Aoki et al., FLAG working group, arXiv:1310.8555 [hep-lat] (2013).
 
 \bibitem{PERIS}E. de Rafael, {\it Nucl. Phys. Proc. Suppl.} {\bf 96} (2001) 316; S. Peris, B. Phily and E. de Rafael, {\it Phys. Rev. Lett.} {\bf 86} (2001) 14.
 
 \bibitem{SNREV14}See e.g. S. Narison, {\it Nucl. Phys. Proc. Supp.} {\bf 258-259} (2015) 189. 
 
 \bibitem{STEVENSON}P.M. Stevenson, {\it Nucl.Phys.} {\bf B868} (2013) 38;
S. J. Brodsky, G. P. Lepage and P. B. Mackenzie, {\it Phys. Rev.} {\bf D28} (1983) 228; X.-G. Wu et al.,  arXiv:1405.3196 [hep-ph] (2014); A.L. Kataev and S.V. Mikhailov, {\it Phys.\ Rev.} {\bf D91} (2015) 1,  014007; J. -L Kneur and A. Neveu, {\it Phys.Rev.} {\bf D88} (2013) 074025.

 \bibitem{BEC} S. Bekavac et al.,    {\it Nucl. Phys. }  {\bf B833} (2010), 46 and references therein.


\bibitem{PIVOV}P. Gelhausen et al., {\it Phys. Rev.} {\bf D88} (2013) 014015.

 \bibitem{DAVIES}  B. Colquhoun et al., {\it Phys. Rev.} {\bf D91} (2015) 114509; private communication from Christine Davies.

    \bibitem{LUCHA2}  W. Lucha, D. Melikhov and S. Simula, 
    {\it Phys. Lett.} {\bf B735} (2014) 12; 
  W. Lucha, D. Melikhov and S. Simula, {\it Phys. Rev.} {\bf D91} (2015) 116009; private communication from Dimitri  Melikhov.
    
  \bibitem{DAI} Y.B Dai et al., {\it Eur. Phys. J.} {\bf C55} (2008) 249.

\bibitem{COLA2}P. Colangelo et al., {\it Phys. Lett.} {\bf B269} (1991) 201.

\bibitem{WANG} Z.-G Wang , {\it Eur. Phys. J.} {\bf C75} (2015) 9, 427; private communication.

\bibitem{JUGEAU}F. Jugeau et al., {\it Phys. Rev.} {\bf D72}, 094010 (2005).
 
 \bibitem{HERD}G.Herdoiza, C. McNeile and C. Michael, {\it Phys. Rev.} {\bf D74} (2006) 014510.
 

  \bibitem{BAGAN}E. Bagan et al., {\it Z. Phys.} {\bf C64} (1994) 57.
  
  \bibitem{CHABAB}M. Chabab, {\it Phys. Lett.}  {\bf B325} (1994)  205;
P. Colangelo, G. Nardulli and N. Paver, {\it Z. Phys.} {\bf C57} (1993) 43.

 \bibitem{ETM14} N. Carrasco et al., {\it Phys. Rev.} {\bf D91} (2015) 054507.

\bibitem{MILC}  A. Bazavov et al., {\it Phys. Rev.} {\bf D90} (2014) 074509.

\bibitem{LUCHA1}  W. Lucha, D. Melikhov and S. Simula, {\it Phys. Lett.} {\bf B701} (2011) 82.

\bibitem{PENA}J. Bordes, J. Penarrocha and K. Schilcher, {\it JHEP} {\bf 0511} (2005) 014; ibid, {\it JHEP} {\bf 0412} (2004) 064.

    


 

 \bibitem{ETM13} A. Bussone et al., arXiv:1411.5566 [hep-lat] (2014); 
 
 \bibitem{HPQCD13}R. Dowdall et al., {\it Phys. Rev. Lett.} {\bf 110} (2013) 222003.
 
 \bibitem{PENA2}M. J Baker  et al., {\it JHEP} {\bf 1407} (2014) 32.

    \bibitem{LUCHA3}  W. Lucha, D. Melikhov and S. Simula, {\it Phys. Rev.} {\bf D88} (2013) 05601.







 \bibitem{LATTFBC}C. McNeile et al., {\it Phys. Rev.} {\bf D86} (2012) 074503.
         
\bibitem{DOM} M. Baker et al., arXiv:1310.0941 [hep-ph] (2013). 

\bibitem{DAVIES2}G. Donald et al., {\it Phys. Rev. Letters} {\bf 112} (2014) 212002.

\bibitem{HWANG} C.-W Hwang, {\it Phys. Rev.} {\bf D81} (2010) 054022 and references therein.


\bibitem{HQET} M.B. Voloshin and M.A. Shifman, {\it Sov. J. Nucl. Phys.} {\bf 45} (1987) 292; H.D. Politzer and M.B. Wise, {\it Phys. Lett.}  {\bf B206} (1988)  504,681; F. Hussain et al.,  {\it Phys. Lett.} {\bf B249} (1990)  295;  E. Eichten,   {\it Nucl. Phys. Proc. Suppl.} {\bf 20} (1991) 475.

\bibitem{SNIMET} S. Narison, {\it Z. Phys.} {\bf C55} (1992) 671; S. Narison, {\it Phys. Lett.}  {\bf B279} (1992)  137;  S. Narison, {\it Phys. Lett.}  {\bf B308} (1993)  365.


\end{thebibliography}
\end{document}